\documentstyle[aps,epsfig,floats,amssymb]{revtex}
\draft

\begin{document}
\twocolumn[\hsize\textwidth\columnwidth\hsize\csname
@twocolumnfalse\endcsname

\title{Gravity from Spinors}

\author{C. Wetterich}

\address{
Institut f{\"u}r Theoretische Physik,
Philosophenweg 16, 69120 Heidelberg, Germany}

\maketitle

\begin{abstract}
We investigate a possible unified theory of all interactions which is based
only on fundamental spinor fields. The vielbein and metric arise as composite objects.
The effective quantum gravitational theory can lead to
a modification of Einstein's equations due to
the lack of local Lorentz-symmetry. We explore the generalized gravity with global
instead of local Lorentz symmetry in first order of
a systematic derivative expansion. At this level diffeomorphisms and global
Lorentz symmetry allow for two new invariants in the gravitational
effective action. The one which arises in the one loop
approximation to spinor gravity is consistent with all present tests of general relativity
and cosmology. This shows that local
Lorentz symmetry is tested only very partially by present observations.
In contrast, the second possible new coupling is severely
restricted by present solar system observations.
\end{abstract}
\pacs{PACS numbers: 12.10.-g; 04.20.Cv; 11.10.Kk  \hfill HD-THEP-03-32}

 ]

\section{Introduction}
\label{introduction}
Can a fundamental theory of all interactions be based only on spinors? The fermions as
crucial constituents of matter are indeed described by spinor fields. In contrast, the
interactions are mediated by bosons which do not have the transformation properties of
spinors. Any realistic spinor theory has therefore to account for the bosons as bound states.
In principle, this poses no problem since bosons may be composed of an even number of
fermions. In a fundamental theory, however, we need bosons with very particular properties:
the graviton is connected to the symmetry of general coordinate transformations
(diffeomorphisms) and the gauge interactions are mediated by gauge bosons with spin one.
Furthermore, scalar fields are needed in order to achieve the spontaneous breaking of the
electroweak symmetry and possibly also extended symmetries like grand unified gauge
symmetries. This raises \footnote{The old question of a fundamental spinor theory has been
discussed in different contexts by De Broglie, Heisenberg and many others. Here we address
more specifically the problem if gravity can arise from spinors.}
a first challenge: can gravity arise from a spinor field theory?

Several proposals in this direction have discussed ``pregeometry'' \cite{Ak} or
``metric from matter'' \cite{Av}, inspired by the observation that the matter fluctuations
in a gravitational background field can generate a kinetic term for the graviton
\cite{BGR}. While the introduction of a diffeomorphism invariant action for the spinors
is rather straightforward \cite{Ak}, the arguments presented in favor of local Lorentz
symmetry are less convincing. The main obstacle is the absence of an object transforming
as a spin connection that could be constructed as a polynomial of the spinor fields
\footnote{Elements of the Grassmann algebra are polynomials in the spinor fields and can
be classified according to their rank. Local Lorentz transformations do not change the rank
of an object. A covariant derivative needs a spin connection which must have rank 0.
However, no object of rank 0 with the requested inhomogeneous transformation properties
can be constructed from spinor polynomials. Neither the limiting process proposed in
\cite{Ak} nor the covariant derivative suggested in \cite{Av} have a well defined meaning
as polynomials in the spinor fields. In particular, we note that the inverse of the
``Grassmann matrix'' $M_{\alpha\beta}=\psi_\alpha\bar{\psi}_{\beta}$ does not exist for
Grassmann variables with ``spinor index'' $\alpha,\beta$. We note, however, the construction
of a supersymmetric action with local Lorentz symmetry in a setting with nonlinear fields,
including additional bosonic fields \cite{LU}. Other approaches integrate out a bosonic
connection \cite{HSS}.}. Concentrating on a well defined spinor action as a polynomial
in the fermionic Grassmann variables the models that have been proposed so far exhibit only
global instead of local Lorentz invariance \cite{GLo}. Only very recently\footnote{The
present work was performed before this finding.} a locally Lorentz invariant polynomial
spinor action has been found \cite{6AA}.

In this paper we explore the alternative of a spinor action that only respects the global
and not the local Lorentz transformations. Then also the gravitational theory for the
vielbein which emerges in this setting will only exhibit a global Lorentz symmetry.
The quantum fluctuations will lead
to a theory with a massless graviton bound state as well as further massless
bosonic excitations which are responsible for a particular form of torsion.
We will see that local Lorentz symmetry is actually not required by observation.
A new invariant, generated by one loop spinor gravity and violating local Lorentz
symmetry, is compatible with all present tests of general relativity.

Once the graviton can be associated to a bound state of fermions,
the explanation of the other bosonic degrees of freedom could follow a well established road.
A higher dimensional gravity theory can induce four dimensional gauge interactions by
``spontaneous compactification'' \cite{KK}, \cite{Wi}.
The gauge symmetries are then related to the
isometries of ``internal space''. The number of generations of massless or light fermions is
connected to the ``chirality index'' \cite{CWF} which depends on the topology and
symmetries of internal space. A non-vanishing index requires a vanishing higher dimensional
mass term for the fermions \cite{CWMS} and we therefore need an effective higher
dimensional theory with massless fermions and gravitons. Both the ``constituents'' and
``bound states'' need to be massless! Finally, the higher dimensional metric also contains
four dimensional scalar fields with the properties required for
spontaneous symmetry breaking. Rather
realistic models with the gauge interactions of the standard model, three chiral generations
of quarks and leptons, spontaneous symmetry breaking and an interesting hierarchical pattern
of fermion masses and mixings have been proposed \cite{CWFA} based on 18 dimensional gravity
coupled to a Majorana-Weyl spinor.

The requirement of a local polynomial spinor action which is invariant under general
coordinate transformations leads to the proposal \cite{HW1}
of ``spinor gravity'' as a possible fundamental theory of all interactions. In the
present accompanying paper we elaborate on this proposal and put it into a somewhat more
general context. In particular, we discuss here the role of the additional
``gravitational'' degrees of freedom which are
due to the lack of local Lorentz symmetry. These
massless excitations, which have not been discussed previously, lead to possible
modifications of Einstein's gravity on macroscopic scales. Comparison with the present
status of observations will reveal that the usual assumption of local Lorentz symmetry is
actually very poorly tested. It is possibly to add to Einstein's action a new invariant
which preserves global but not local Lorentz symmetry and which is nevertheless consistent
with all present observations. In this case, the new massless gravitational degrees of
freedom couple only to macroscopic spin. A second paper \footnote{The content of this
second paper was contained in the first version of the present paper.} \cite{SG2} will
discuss the nonlinear geometrical structures of our setting.

Fermion bilinears transforming as vector fields under general coordinate
transformations can be obtained from derivatives, $\tilde{E}^m_\mu=i\bar{\psi}\gamma^m
\partial_\mu\psi/2+h.c$. Here $\psi(x)$ denotes Grassmann variables in the spinor
representation of the $d$-dimensional Lorentz group and we have introduced the associated
Dirac matrices $\gamma^m$ such that $\tilde{E}^m_\mu$ is a vector with respect to global
Lorentz rotations. \footnote{In three dimensions, a similar object can be used to
characterize the order parameter of liquid $He^3$ \cite{He}.}
From $\tilde{E}^m_\mu$ we can construct a composite operator
with the transformation properties of the metric, $\tilde{E}^m_\mu\tilde{E}_{\nu m}$.
However, the action has to be a polynomial in the spinors and no object transforming as
the inverse metric can be used in order to contract the ``lower world indices'' connected to
the derivatives. The only possible choice for a diffeomorphism invariant action therefore
contracts $d$ derivatives with the totally antisymmetric tensor $\epsilon^{\mu_1\dots\mu_d}$.
Invariance under global Lorentz  rotations can be achieved similarly by contraction with
$\epsilon_{m_1\dots m_d}$. In consequence, it is indeed possible to construct an invariant
action as a local polynomial in the spinor fields and their derivatives
\begin{eqnarray}\label{AA0}
S_E&=&\alpha\int d^dx\det\big(\tilde{E}^m_\mu(x)\big)~,\nonumber\\
\tilde{E}^m_\mu&=&\frac{i}{2}\bar{\psi}\gamma^m\partial_\mu\psi+h.c.~.
\end{eqnarray}
We emphasize that we have no spin connection at our disposal. Therefore the bilinear
$\tilde{E}^m_\mu$ does {\em not} transform as a tensor under {\em local} Lorentz
transformations. Instead, its transformation property is characterized by an additional
inhomogeneous piece involving the derivative of the Lorentz transformation parameter.

In consequence, the action $S_E$ is invariant under {\em global} but not {\em local} Lorentz
transformations. This is an important difference as compared to the standard formulation
of gravity (``Einstein gravity'').
We will explore both the phenomenological and conceptual aspects of this difference.
Actually, the action (\ref{AA0}) is not the only invariant
with diffeomorphism and global Lorentz symmetry - other invariants are discussed in
\cite{SG2}. We will see that the lack of local Lorentz symmetry leads to a generalized
version of gravity.

Within ``spinor gravity'' the
``global vielbein'' $E^m_\mu(x)$ can be associated to the expectation
value of the fermion bilinear $\tilde{E}^m_\mu(x)$. As usual
the metric obtains then by contraction
with the invariant tensor $\eta_{mn}$ which lowers the Lorentz indices
\begin{equation}\label{AA1}
E^m_\mu(x)=\langle\tilde{E}^m_\mu(x)\rangle~,~g_{\mu\nu}(x)=E^m_\mu(x)E_{\nu m}(x).
\end{equation}
On the level of the composite bosonic fields $E^m_\mu$ and $g_{\mu\nu}$ the inverse vielbein
and metric $E^\mu_m(x)~,~g^{\mu\nu}(x)$ are well defined provided $E=\det(E^m_\mu)\neq 0$.
The field equations for the vielbein and metric can, at least in principle, be computed
from eq. (\ref{AA0}) plus an appropriate regularization of the functional measure.
This approach realizes the general idea that both geometry
and topology can be associated to the
properties of appropriate correlation functions \cite{GenG} - in the present case the
two point functions for spinors.

Due to the lack of local Lorentz symmetry the global vielbein contains additional degrees of
freedom that are not described by the metric. Correspondingly,
the effective theory of gravity will also
exhibit new invariants not present in Einstein gravity. These invariants
are consistent with global but not local Lorentz symmetry.
Indeed, we may use a nonlinear field decomposition $E^m_\mu(x)=e^m_\mu(x)H_m^{\ n}(x)$ where
$e^m_\mu$ describes the usual ``local vielbein'' and $H_m^{\ n}$ the additional degrees of
freedom. These additional degrees of freedom are massless Goldstone-boson-like
excitations due to the spontaneous breaking of a global symmetry. In Einstein gravity,
$H_m^{\ n}$ would be the gauge degrees of freedom of the local Lorentz transformations and
therefore drop out of any invariant action. In contrast, the generalized gravity discussed
here will lead to new propagating massless gravitational degrees of freedom.

Indeed, the
kinetic terms for $H_m^{\ n}$ can be inferred from the most general effective action
for the vielbein which contains
two derivatives and is invariant under diffeomorphisms and global Lorentz
transformations
\begin{eqnarray}\label{AA2}
\Gamma_{(2)}&=&\frac{\mu}{2}\int d^dxE\Big\{-R+\tau_A[D^\mu
E^\nu_mD_\mu E^m_\nu\nonumber\\
&&-2D^\mu E^\nu_m D_\nu E^m_\mu]
+\beta_A D_\mu E^\mu_m D^\nu E^m_\nu\Big\}.
\end{eqnarray}
Here the curvature scalar $R$ is constructed from the metric $g_{\mu\nu}$ which is also used
to lower and raise world indices in the usual way. The covariant derivative $D_\mu$
contains the connection $\Gamma_{\mu\nu}^{\ \ \lambda}$ constructed from $g_{\mu\nu}$ but
no spin connection. Due to the missing spin connection the last two terms
$\sim\tau_A,\beta_A$ are invariant under global but not local Lorentz transformations. They
induce the kinetic term for $H_m^{\ n}$. The effective action (\ref{AA2}), together with a
``cosmological constant'' term $\sim\int d^dxE$, constitutes the first order in a systematic
derivative expansion. In the one loop approximation to spinor gravity
one finds $\beta_A=0$.

In this paper we discuss the viability of generalized gravity \cite{CWGG} in a setting
with only global Lorentz
symmetry. For this purpose we analyze the consequences of the effective action (\ref{AA2})
in four dimensions. In complete analogy to Einstein gravity we discuss the solutions of the
field equations derived from the effective action (\ref{AA2}) in presence of suitable
sources associated to an energy momentum tensor. In principle, the energy momentum 
tensor contains an antisymmetric part $T^{\mu\nu}_A$ which reflects the presence of anomalous 
spin interactions for the fermions. These effects are, however, much too small to be observable. 
Then $T^{\mu\nu}_A$ can be neglected and test particles couple to the metric
in the usual way. We find that for $\beta_A=0$ neither Newtonian gravity nor the isotropic
Schwarzschild or the cosmological Friedman solutions are modified. This also holds
for the emission, propagation and detection of gravitational waves and for all tests of general 
relativity in post-Newtonian gravity. For vanishing 
$\beta_A$ our generalized gravity is therefore consistent with all present observations of
general relativity! We conclude that a violation of local Lorentz symmetry by the invariant
$\sim \tau_A$ in eq. (\ref{AA2}) remains unconstrained experimentally.
On the other hand, for $\beta_A\neq 0$
we find a modification of the Schwarzschild
solution similar to a Jordan-Brans-Dicke theory \cite{JBD}: whereas $g_{00}=-B(r)$ behaves as
usual as $B(r)=1-r_s/r$ (with $r_s$ the Schwarzschild radius), one obtains
$g_{rr}=A(r)=(1-\gamma r_s/r)^{-1}$ where $\gamma \approx 1+\beta_A$. This imposes a severe
bound $|\beta_A|\stackrel{<}{\sim}5\cdot10^{-5}$ \cite{Ber}.
In view of this bound the modifications of
cosmology are too small to be presently observable \footnote{Spinor gravity may lead to
other long range degrees of freedom not contained in $E^m_\mu$. These could lead to
interesting modifications of gravity like quintessence \cite{Q,CQ}.}.

This paper is organized as follows. In sect. \ref{invariantspinor} we recapitulate the transformation
properties of spinor fields and bilinears and the construction of the
polynomial  action (\ref{AA0}). The effective
bosonic action for fermion bilinears is formulated in sect. \ref{bosoniceffective}.
This setting describes our version of quantum gravity.
In sect. \ref{symmetriesand} we start a general discussion of gravity theories with only global instead of
local Lorentz symmetry. There we classify the possible invariants with up to two derivatives
and formulate the effective action in first order in a systematic derivative expansion.
In addition to the terms present in Einstein gravity it contains the two invariants
(\ref{AA2}) with dimensionless coefficients $\tau_A$ and $\beta_A$.
The corresponding generalized gravitational
field equations are derived in sect. \ref{curvaturescalar}.
In sects. \ref{linearizedgravity} and \ref{linearizedfield}
we discuss the linear approximation
to the field equations. Beyond the graviton of Einstein gravity the spectrum of excitations
contains a new set of massless fields described by an antisymmetric tensor field
$c_{\mu\nu}$. However, this field does not couple to the symmetric part of the energy
momentum tensor but rather to the antisymmetric part which reflects the internal degrees of
freedom of the spinors. We show in sect. VIII that the 
new interactions mediated by the exchange of $c_{\mu\nu}$ play
no macroscopic role and do not affect the observational effects of
linear gravity. We also establish that the invariant $\sim\tau_A$ is compatible with all 
tests of general relativity in first nonleading order in post-Newtonian gravity.

In the linear approximation one finds for $\beta_A\neq 0$ also an
additional massless vector field $w_\mu$. Again, it couples only to the antisymmetric
part of the energy momentum tensor. More important, a nonvanishing coupling $\beta_A$ modifies
also the linearized equation for the degrees of freedom contained in the metric - more
precisely the coupling of the ``conformal factor'' $\sigma$. In the Newtonian approximation
this effect only renormalizes Newton's constant. Beyond Newtonian gravity $\beta_A\neq 0$
affects the tests of general relativity.

Going beyond the linear and post-Newtonian approximations,
we discuss the modifications of the general isotropic static solution and
the homogeneous isotropic cosmological solution for the full field equations.
In sect. \ref{generalisotropic} we present the generalization
of the isotropic static metric for a gravity theory with only global Lorentz symmetry. The
corresponding modification of the Schwarzschild solution for $\beta_A\neq 0$ is discussed in
sect. \ref{modificationofthe}.
In sect. \ref{homogeneousisotropic} we turn to the
most general homogeneous and isotropic cosmological
solution within our setting of generalized gravity. We find that the solutions of Einstein
gravity remain also solutions of generalized gravity as long as $\beta_A=0$. For
$\beta_A\neq 0$ one finds a difference in the value of the Planck mass appearing in the
cosmological equations as compared to the one inferred from the Newtonian approximation.
In view of the solar system bounds on $|\beta_A|$ this effect is too small in order to
be presently observable. For small enough nonvanishing 
$|\beta_A|$ the generalized gravity with only global Lorentz symmetry obeys 
all present tests of general relativity. Similar to the Brans-Dicke theory the model with 
small nonzero $\beta_A$ can be used to quantify the experimental precision of general relativity. 
It is therefore interesting in its own right and merits further quantitative studies in the future  
- even though spinor gravity may finally result in $\beta_A=0$. 

In sects. \ref{partialbosonization} and \ref{generalizeddirac} we make a first attempt
to compute the bosonic effective action for spinor gravity. 
For this purpose we express the fermionic
functional integral in terms of a ``partially bosonized'' functional integral.
We briefly explore the classical approximation to the field equations. In sect.
\ref{generalizeddirac} we discuss
the generalized Dirac operator in an arbitrary ``background geometry'' $E^m_\mu$. This
defines the one loop approximation to spinor gravity. One loop spinor gravity only
involves the new invariant $\sim\tau_A$ that is not restricted by observation.
In our conclusions in sect. \ref{conculsions} we finally
discuss the prospects for spinor gravity as a
candidate for a unified theory of all interactions.

\section{Invariant spinor action}
\label{invariantspinor}
Our basic entities are spinor fields $\psi(x)$ which are represented by anticommuting
Grassmann variables and transform as irreducible spinor representations under the
$d$-dimensional Lorentz group $SO (1,d-1)$
\begin{equation}\label{1}
\delta_{{\cal L}}\psi~=~-\frac{1}{2}\epsilon_{mn}
\Sigma^{mn}\psi~,~\Sigma^{mn}=-\frac{1}{4}[\gamma^m,\gamma^n].
\end{equation}
Here the Dirac matrices obey $\{\gamma^m,\gamma^n\}=2\eta^{mn}$, and Lorentz indices
are raised and lowered by $\eta^{mn}=\eta_{mn}=~diag~(-1,+1,\dots,+1)$. Under $d$-dimensional
general coordinate transformations the spinor fields transform as scalars
\begin{equation}\label{2}
\delta_\xi\psi=-\xi^\nu\partial_\nu\psi
\end{equation}
such that $\partial_\mu\psi$ is a vector. Similarly, the spinor fields $\bar{\psi}(x)$
transforms as
\begin{equation}\label{3}
\delta_{{\cal L}}\bar{\psi}=\frac{1}{2}\bar{\psi}\epsilon_{mn}\Sigma^{mn}~,~
\delta_\xi\bar{\psi}=-\xi^\nu\partial_\nu\bar{\psi}.
\end{equation}
For Majorana spinors in $d=0,1,2,3,4~mod~8$ one has $\bar{\psi}=\psi^TC$ where $C$
obeys \footnote{For details see \cite{CWMS}} $(\Sigma^T)^{mn}=-C\Sigma^{mn}C^{-1}$.
Otherwise $\bar{\psi}$ may be considered as an independent spinor, with an involutive
mapping between $\psi$ and $\bar{\psi}$ associated to complex conjugation in spinor space.
In even dimensions the irreducible spinors are Weyl spinors obeying
$\bar{\gamma}\psi=\psi$ with $\bar{\gamma}=\eta\gamma^0\dots\gamma^{d-1},
\eta^2=(-1)^{d/2-1},\bar{\gamma}^2=1,\bar{\gamma}^\dagger=\bar{\gamma}$. Majorana-Weyl
spinors exist for $d=2~mod~8$.

We want to construct an action that is a polynomial in $\psi,\bar{\psi}$ and invariant
under global Lorentz-transformations and general coordinate tansformations. Our basic
building block is a spinor bilinear\footnote{In \cite{SG2} we generalize this construction 
to an bilinear $\tilde{E}^m_\mu=i\bar{\psi}\gamma^m\partial_\mu\psi$ that is not 
necessarily hermitean.}
\begin{equation}\label{3AA}
\tilde{E}^m_\mu=\frac{i}{2}(\bar{\psi}\gamma^m\partial_\mu
\psi-\partial_\mu\bar{\psi}\gamma^m\psi).
\end{equation}
It transforms as a vector under general coordinate
transformations
\begin{equation}\label{4}
\delta_\xi\tilde E^m_\mu=-
\partial_\mu\xi^\nu\tilde E^m_\nu-\xi^\nu\partial_\nu\tilde E^m_\mu,
\end{equation}
and as a vector under global Lorentz rotations
\begin{equation}\label{4a}
\delta_{\cal L}\tilde E^m_\mu=\epsilon^m_{\ n}\tilde E^n_\mu.
\end{equation}
For irreducible spinors in $d=2,3,9~mod~8$ one has $\bar{\psi}\gamma^m\psi=0$ such that
$\tilde{E}^m_\mu=i\bar{\psi}\gamma^m\partial_\mu\psi$.
From $\tilde E^m_\mu$ we can easily construct a composite field transforming like the metric
\begin{equation}\label{5}
\tilde g_{\mu\nu}=\tilde E^m_\mu\tilde E^n_\nu\eta_{mn}.
\end{equation}

However, no object transforming as the inverse metric can be constructed as a polynomial
in the spinor fields. The spinor polynomials contain only ``lower world indices'' $\mu,\nu$
which are induced by derivatives. The only possible coordinate invariant polynomial must
therefore involve precisely $d$ derivatives, contracted with the totally antisymmetric
$\epsilon$-tensor. In particular, the scalar density $\tilde E=\det(\tilde E^m_\mu)$ can
be written as a spinor polynomial
\begin{equation}\label{6}
\tilde E=\frac{1}{d!}\epsilon^{\mu_1\dots\mu_d}
\epsilon_{m_1\dots m_d}
\tilde E^{m_1}_{\mu_1}\dots
\tilde E^{m_d}_{\mu_d}=\det(\tilde{E}^m_\mu).
\end{equation}
Therefore a possible invariant action reads
\begin{equation}\label{7}
S_E=\alpha\int d^dx\tilde E.
\end{equation}
It involves $d$ derivatives and $2d$ powers of $\psi$. We note that the ways to construct
invariants are restricted by the absence of objects transforming as the inverse metric
or the inverse vielbein. All invariants contain $\epsilon^{\mu_1\dots\mu_d}$ where the
indices $\mu_1\dots\mu_d$ have to be contracted with derivatives. On the other hand, the
construction of invariants with respect to the global Lorentz symmetry is not unique
\cite{SG2} since we have the invariant tensor $\eta_{mn}$ and spinor bilinears not involving
derivatives at our disposal. 

With respect to local
Lorentz transformations $\delta_{\cal L}\tilde{E}^m_\mu$ acquires additional inhomogeneous
pieces. In fact, if $\epsilon_{mn}(x)$ depends on the spacetime coordinate one has
\begin{eqnarray}\label{12AA}
\delta_{\cal L}\tilde{E}^m_\mu&=&\epsilon^m_{\ n}\tilde{E}^n_\mu+
\left(\frac{i}{8}\bar{\psi}\gamma^{[mnp]}\psi\partial_\mu\epsilon_{np}\right.\nonumber\\
&&\left.+\frac{i}{4}\bar{\psi}\gamma^n\psi\partial^\mu\epsilon^m\ _n+h.c.\right)
\end{eqnarray}
with $\gamma^{[mnp]}$ the totally antisymmetrized product of three $\gamma^m$-matrices, 
$\gamma^{[mnp]}=\frac{1}{6}(\gamma^m\gamma^n\gamma^p-\gamma^m\gamma^p\gamma^n+\dots)$. 
The piece $\sim\bar{\psi}\gamma^n\psi$ drops out - recall that for Majorana spinors in 
$d=2,3,9~mod~8$ the antisymmetry under the exchange of Grassmann variables implies 
$\bar{\psi}\gamma^n\psi=0$ \cite{CWMS}. The piece $\sim\bar{\psi}\gamma^{[mnp]}\psi$ 
remains, however. In consequence,
the action $S_E$ is only invariant under {\it global} Lorentz rotations, but not {\it local}
Lorentz rotations. Two spinor configurations related to each other by a {\em local} Lorentz 
transformation are not equivalent to each other\footnote{On the level of $\tilde{E}^m_\mu$ one may 
formualte a ``new'' local Lorentz transformation by using the transformation rule 
(\ref{4a}) instead of (\ref{12AA}) such that $S_E$ is invariant. However, this transformation 
cannot be formulated on the level of $\psi$. Since the transformtion of the functional measure 
is not defined the effective gravitational action will not obey such a symmetry.}. 

\section{Bosonic effective action}
\label{bosoniceffective}
In order to construct the quantum effective action for our model with classical action
(\ref{7}) we introduce
fermionic sources $\bar\eta$ and
bosonic sources $J^\mu_m$. The fermionic sources are Grassmann variables transforming as
\begin{equation}\label{10}
\delta_{\cal L}\bar\eta=\frac{1}{2}\bar\eta\epsilon_{mn}\Sigma^{mn}~,~
\delta_\xi\bar\eta=-\xi^\nu\partial_\nu\bar\eta-(\partial_\nu\xi^\nu)\bar\eta
\end{equation}
such that $S_\eta=-\int d^dx\bar\eta\psi$ is invariant. The bosonic sources
multiply the fermion bilinear $\tilde E^m_\mu$. With the vector density $J^\mu_m$
transforming as
\begin{eqnarray}\label{11}
\delta_{\cal L}J^\mu_n&=&-J^\mu_m\epsilon^m_{\ n}\nonumber\\
\delta_\xi J^\mu_m&=&-\partial_\nu(\xi^\nu J^\mu_m)+\partial_\nu\xi^\mu J^\nu_m
\end{eqnarray}
the source term
\begin{equation}\label{14XX}
S_J=-\int d^dxJ^\mu_m\tilde E^m_\mu
\end{equation}
is again invariant. The generating
functional \footnote{For $d=5,6,7~mod~8$ we have to use a spinor $\bar\psi$ not
related to $\psi$ by $\bar\psi=\psi^TC$. One should therefore use sources $\bar\eta$ and
$\eta$ multiplying $\psi$ and $\bar\psi$ and a functional measure involving $\psi$ and
$\bar\psi$.}
\begin{equation}\label{12}
W[\bar\eta,J]=\ln Z[\bar\eta,J]=\ln\int D\psi\exp\Big\{-(S+S_\eta+S_J)\Big\}
\end{equation}
is therefore an invariant functional of $\bar\eta$ and $J$ provided that the functional
measure $\int D\psi$ is free of anomalies. The ``vielbein'' is now defined as the expectation
value of $\tilde E^m_\mu$
\begin{equation}\label{13}
\frac{\delta W}{\delta J^\mu_m}=E^m_\mu=
\langle \tilde{E}^m_\mu\rangle=
\frac{i}{2}\langle\bar{\psi}\gamma^m\partial_\mu\psi -\partial_\mu\bar{\psi}\gamma^m
\psi\rangle.
\end{equation}
We use the symbol $E^m_\mu$ instead of the usual $e^m_\mu$ in order to recall that
$E^m_\mu$ does not transform as a vector under {\it local} Lorentz transformations. We omit
here \footnote{The fermionic part of the effective action is discussed in \cite{SG2}.}
the fermionic sources $\bar\eta$ such that $W$ is only a functional of $J$.
Also $E^m_\mu$ depends on $J$. The effective action
\footnote{The sources $J$ can be generalized to multiply arbitrary fermion
bilinears. The ``bosonic effective action'' $\Gamma$ \cite{BEA} contains then all
information about the correlation functions of the system. It can formally be obtained as
the sum over two particle irreducible graphs \cite{BT}.}
for the vielbein is constructed by the
usual Legendre transform
\begin{equation}\label{14}
\Gamma[E^m_\mu]=-W[J^\mu_m]+\int d^dxJ^\mu_mE^m_\mu
\end{equation}
where $J^\mu_m[E^n_\nu]$ obtains by inverting eq. (\ref{13}). It obeys the identity
\begin{equation}\label{15}
\frac{\delta\Gamma}{\delta E^m_\mu}=J^\mu_m.
\end{equation}
If $W$ is an
invariant functional of $J$ we conclude that $\Gamma$ is an invariant functional of the
vielbein $E^m_\mu$.

Eq. (\ref{15}) is the exact gravitational field equation for the quantum field theory
defined by $S_E$ and an appropriate functional measure ${\cal D}\psi$.
In presence of non-gravitational degrees of freedom the ``physical''
source $J^\mu_m$ should be associated to the energy momentum tensor defined by
\begin{equation}\label{18AA}
T^{\mu\nu}=E^{-1}E^{m\mu}J^\nu_m.
\end{equation}
For example, the energy momentum tensor receives
contributions from the spinor fields as well as other possible bosonic composite
fields beyond the vielbein. If the four dimensional effective action is obtained
by dimensional reduction from a higher
dimensional theory, the source $J^\mu_m$ also accounts for the gauge bosons and
scalars which arise from the higher dimensional bosonic fields. If we collect all
contributions to the effective action involving fields other than the vielbein
in $\Gamma'$ we can formally write $J^\mu_m=-\langle\delta\Gamma'/\delta E^m_\mu\rangle$,
where the bracket indicates that $J^\mu_m$ has to be evaluated for the given physical state
\footnote{For the formal construction one introduces the sources $J^\mu_m$ only as technical
devices and puts them to zero at the end of the computation. However, one has to compute
$\Gamma+\Gamma'$ with field equation
$\delta\Gamma/\delta E^\mu_m+\delta\Gamma'/\delta E^\mu_m=0$. The piece from the variation
of $\Gamma'$ can then be reinterpreted as a nonvanishing ``physical source''.}. For
example, $J^\mu_m$  may account for the presence of a macroscopic massive object or for
a relativistic plasma in cosmology. This setting is completely analogous to the
treatment of standard gravity. As usual in gravity we define \footnote{In principle,
$g_{\mu\nu}$ differs from the fermionic four point function
$\langle\tilde{g}_{\mu\nu}\rangle$ (cf. eq. (\ref{5})).} the metric by
\begin{equation}\label{18AB}
g_{\mu\nu}=E^m_\mu E_{\nu m}.
\end{equation}

We next show that for most practical purposes $T^{\mu\nu}$ can be identified with the
usual energy momentum tensor. Consider first the effective action
$\Gamma'_0$ for fields with trivial Lorentz-transformation properties as scalars or
gauge bosons (i.e. fields carrying only ``world indices'' $\mu,\nu$ and no spinor or
Lorentz index). Then the dependence of $\Gamma'$ on the gravitational degrees of
freedom arises only via the metric
\begin{equation}\label{18AC}
\Gamma'_0[E^m_\mu]=
\Gamma'_0\Big[g_{\nu\rho}[E^m_\mu]\Big].
\end{equation}
In particular, this holds for structureless point particles. In standard gravity the energy
momentum tensor $T^{\mu\nu}_{(g)}$ is defined as
\begin{equation}\label{18AD}
T^{\mu\nu}_{(g)}=-\frac{2}{\sqrt{g}}\frac{\delta\Gamma'_0}{\delta g_{\mu\nu}}.
\end{equation}
Using the definition (\ref{18AA}) we find
\begin{equation}\label{18AE}
T^{\mu\nu}=-E^{-1}E^{m\mu}\frac{\delta\Gamma'_0}{\delta g_{\rho\sigma}}
\frac{\delta g_{\rho\sigma}}{\delta E^m_\nu}=T_{(g)}^{\mu\nu}
\end{equation}
and conclude that $T^{\mu\nu}$ is symmetric and coincides indeed with the standard
energy momentum tensor.

One may object that the situation is different for fields
with non-trivial Lorentz-transformation properties as, for example, spinors. Then
the gravitational couplings can typically not be written only in terms of the metric but
involve explicitely the vielbein. We will discuss in sect. VIII that the fermion
contribution to the energy
momentum tensor involves an antisymmetric part \cite{HW1,SG2} proportional to the spin.
Nevertheless, for standard macroscopic gravitational sources the spin averages out and
only the symmetric part of $T^{\mu\nu}$ needs to be retained \footnote{Note that orbital
angular momentum does not contribute to the antisymmetric part of $T^{\mu\nu}$.}. For stars,
dust and radiation we can write the gravitational field equation in terms of the usual
energy momentum tensor
\begin{equation}\label{18AF}
\frac{\delta\Gamma}{\delta E^m_\mu}=EE_{\rho m}T_{(g)}^{\rho\mu}.
\end{equation}

Similar considerations hold for test particles used to probe the gravitational fields
generated by other bodies. The action for photons depends only on the metric - their
trajectory can therefore be computed as usual once $g_{\mu\nu}$ is known. Similarly,
macroscopic test particles follow the geodesics defined by $g_{\mu\nu}$.
Throughout this paper we will assume that gravity is tested by point particles or light. 
Testable differences between our setting and Einstein's gravity can
therefore only result from possible differences of the solutions of the field equation
(\ref{18AF}) as compared to the Einstein equations.

At this point we would like to stress that a successful computation of
$\Gamma[E^m_\mu]$ (together with $T^{\mu\nu}$ or $\Gamma'$) is equivalent to a well
defined theory of quantum gravity. The gravitational field equation
(\ref{15}) includes all quantum fluctuations. Also the motion of test particles can
directly be inferred from $\Gamma'$. The difficult part is, of course, the computation
of $\Gamma$. In particular, this requires a well defined functional measure
${\cal D}\psi$ which preserves diffeomorphisms and global Lorentz symmetry.

\section{Symmetries and invariants}
\label{symmetriesand}
We will make a first attempt to a very approximate computation of $\Gamma[E^m_\mu]$ in
sects. \ref{partialbosonization}, \ref{generalizeddirac}. Before, we want to exploit the
general structure of the effective action, in particular the symmetries. This
will allow us a first judgment if a theory with only global Lorentz symmetry
is viable at all. Perhaps surprisingly, we find a new diffeomorphism invariant
involving second derivatives of the vielbein which seems compatible with all present
tests of gravity. This invariant respects global but not local Lorentz symmetry. We
conclude that the local character of the Lorentz symmetry is only very partially
tested - an invariant violating the local symmetry seems to be allowed and remains
essentially unconstrained. On the other hand, we also discuss a second global invariant
which modifies post-Newtonian gravity. Its coefficient is severely constrained.

In sects. \ref{symmetriesand}-\ref{homogeneousisotropic}
we discuss the properties of a generalized version of gravity which features
only global instead of local Lorentz invariance. We discuss the most general setting
consistent with these symmetries. Within spinor gravity this generalizes
the action (\ref{7}) to an arbitrary polynomial action for spinors with invariance
under general coordinate and global Lorentz transformations. (See \cite{SG2} for a
discussion of possible invariants.) Our discussion
will be based purely on symmetry and a derivative expansion of the effective
action. It is therefore more general than the specific one loop approximation
discussed in \cite{HW1}.
The only assumption entering implicitely the following discussion is that the functional
measure preserves diffeomorphism and global Lorentz symmetry, being free of anomalies
\cite{AW}.

In the following we will mainly concentrate on four dimensions, $d=4$. This permits a
direct comparison of the solution of our generalized gravitational field equation
(\ref{18AF}) with observation. Embedding the four dimensional effective theory in a more
fundamental higher dimensional theory we only assume that the ``ground state'' properties of
``internal space'' are consistent with the four dimensional diffeomorphisms and global
Lorentz rotations. Other details are not important for our discussion of the purely
gravitational part. Of course, there could be additional light degrees of freedom
influencing cosmology or the macroscopic laws, like the cosmon of quintessence
\cite{Q}.

Let us therefore discuss the most general
structure of a ``gravitational'' effective action which
involves the vielbein $E^m_\mu$ and is invariant under diffeomorphisms and the global
Lorentz symmetry. The characteristic mass scale will be the Planck mass. For macroscopic
phenomena on length and time scales much larger than the Planck length we can expand
$\Gamma[E^m_\mu]$ in the number of derivatives. For a given number of
derivatives $\Gamma$ can be composed of terms that are each invariant under general coordinate
transformations and global Lorentz rotations.
As compared to Einstein's gravity we will find new invariants which involve the new
``physical'' degrees of freedom in $E^m_\mu$ not described by the metric.

In lowest order in the derivative expansion the unique invariant is
$(E=\det E^m_\mu,g=|\det g_{\mu\nu}|=E^2)$
\begin{equation}\label{S2}
\Gamma_1=\int d^dxE=\pm\int d^dx\sqrt{g}.
\end{equation}
In four dimensions this is a cosmological constant. In the following we
will assume that some mechanism makes the effective cosmologocal constant very small - for
example the dynamical mechanism proposed for quintessence \cite{Q}, \cite{CQ}. We mainly
will discard this term for the following phenomenological discussion.
This is, of course, a highly nontrivial assumption, meaning that
spinor gravity admits an (almost) static solution with large
three-dimensional characteristic length scale (at least the size of the horizon).

For the construction of invariants involving derivatives of $E^m_\mu$
we can employ the antisymmetric tensor
\begin{equation}\label{S3}
\Omega_{\mu\nu}^{\ \ m}=-\frac{1}{2}(\partial_\mu E^m_\nu-\partial_\nu E^m_\mu).
\end{equation}
Let us first look for possible polynomials in $E^m_\mu$ and
$\Omega_{\mu\nu}^{\ \ m}$.
We will concentrate on even dimensions where we need an even power of
$\Omega_{\mu\nu}^{\ \ \ m}$ because of global Lorentz invariance.
Any polynomial invariant with two derivatives must be of the form
\begin{eqnarray}\label{S4}
\Gamma_{2,p}&&=\int d^dx\Omega_{\mu_1\mu_2}^{\ \ \ \ \ n_1}
\Omega_{\mu_3\mu_4}^{\ \ \ \ \ n_2}\nonumber\\
&&E^{m_5}_{\mu_5}\dots E^{m_d}_{\mu_d}\epsilon^{\mu_1\dots\mu_d}
A_{n_1n_2m_5\dots m_d}
\end{eqnarray}
where $A$ should be constructed from $\eta$ and $\epsilon$-tensors and has to be symmetric
in $(n_1,n_2)$ and totally antisymmetric in $(m_5\dots m_d)$. Only for $d=4$ we can take
$A_{n_1n_2}=\eta_{n_1n_2}$ whereas no polynomial two-derivative invariant exists for
$d>4$. (The polynomial invariant generalizing (\ref{S4}) in $d$-dimensions
involves $d/2$ factors of $\Omega$ and therefore $d/2$ derivatives of $E^m_\mu$.) We
observe that $\Gamma_{2,p}$ contains only one $\epsilon$-tensor and therefore
violates parity. We will assume that the gravitational effective action preserves parity,
at least to a very good approximation, and discard the polynomial invariant
(\ref{S4}).

There is, however, no strong reason why the effective action should be a polynomial in
$E^m_\mu$. Whenever $E\neq 0$ we can construct the inverse vielbein
\begin{eqnarray}\label{S5}
E^{\mu_1}_{m_1}&=&\frac{1}{(d-1)!E}\epsilon^{\mu_1\dots\mu_d}\epsilon_{m_1\dots m_d}
E^{m_2}_{\mu_2}\dots E^{m_d}_{\mu_d}\nonumber\\
&=&\frac{1}{E}\frac{\partial E}{\partial E^{m_1}_{\mu_1}}
\end{eqnarray}
which obeys $E^m_\mu E^\nu_m=\delta^\nu_\mu~,~E^\mu_m E^n_\mu=\delta^n_m$.
This allows us to define the inverse metric
\begin{equation}\label{S6a}
g^{\mu\nu}=E^\mu_mE^{m\nu}~,~g^{\mu\nu}g_{\nu\rho}=\delta^\mu_\rho
\end{equation}
which can be used to raise world indices, e.g.
\begin{equation}\label{S7}
\Omega^{\mu\nu}_{\ \ m}=\eta_{mn}g^{\mu\rho}g^{\nu\sigma}\Omega_{\rho\sigma}^{\ \ n}.
\end{equation}
We conclude that non-polynomial invariants exist in arbitrary dimensions.
They are well defined as long as $E\neq 0$ and may become singular in the limit
$E\rightarrow 0$. Within spinor gravity the nonpolynomial invariants are induced by the
fluctuation (or loop) effects \cite{HW1}.

On the level of two derivatives three linearly independent non-polynomial invariants
are given by
\begin{eqnarray}\label{S8}
\Gamma_{2,1}&=&\int d^dxE\Omega^{\ \mu\nu}_{\ \ \ m}\Omega_{\mu\nu}^{\ \ m}\\
\Gamma_{2,2}&=&\frac{1}{2}\int d^dxE(D_\mu E^\mu_m)(D^\nu E^m_\nu)\nonumber\\
\Gamma_{2,3}&=&\frac{1}{4}\int d^dxE(D^\mu E^\nu_m+D^\nu E^\mu_m)
(D_\mu E^m_\nu +D_\nu E^m_\mu).\nonumber
\end{eqnarray}
For the latter two invariants we introduce the covariant derivative
\begin{eqnarray}\label{S10}
D_\mu E^m_\nu&=&\partial_\mu E^m_\nu-\Gamma_{\mu\nu}^{\ \ \lambda} E^m_\lambda~,\nonumber\\
D_\mu E^\nu_m&=&\partial_\mu E^\nu_m+\Gamma_{\mu\lambda}^{\ \ \nu} E^\lambda_m
\end{eqnarray}
where the affine connection involves the inverse metric
\begin{equation}\label{S11}
\Gamma_{\mu\nu}^{\ \ \lambda}=
\frac{1}{2}g^{\lambda\rho}(\partial_\mu g_{\nu\rho}+\partial_\nu
g_{\mu\rho}-\partial_\rho g_{\mu\nu}).
\end{equation}
We emphasize that the covariant derivative acting on $E^m_\mu$ does not contain a spin
connection since $m$ is only a global Lorentz index.
Using the relations (\ref{18AB}), (\ref{S5}) the connection can be expressed in terms of
$E^m_\mu$ in a non-polynomial way.
For $d\neq 4$, or requesting parity invariance, there are no more independent
invariants in this order of the derivative expansion.

The most general invariant bosonic effective action involving up to two derivatives of the
vielbein can therefore be written as a linear combination of $\Gamma_0$ and
$\Gamma_{2,1},\Gamma_{2,2},\Gamma_{2,3}$. As a convenient
parameterization we use
\begin{equation}\label{27NNA}
\Gamma=\epsilon\Gamma_0+\mu(I_1+\tau_A I_2+\beta_A I_3)
\end{equation}
with
\begin{eqnarray}\label{27NNB}
I_1&=&\frac{1}{2}\int d^dxE \{D^\mu E^\nu_mD_\nu E^m_\mu
-D_{\mu}E^\mu_mD^\nu E^m_\nu\}\\
I_2&=&\frac{1}{2}\int d^dxE\{D^\mu E^\nu_mD_\mu E^m_\nu
-2D^\mu E^\nu_mD_\nu E^m_\mu\}\label{27NNC}\\
I_3&=&\frac{1}{2}\int d^dxED_\mu E^\mu_mD^\nu E^m_\nu.\label{27NND}
\end{eqnarray}
We will see that $\mu$ determines the effective Planck mass. This is most apparent
if we rescale $E^m_\mu$ by an arbitrary unit of mass $m$ in order
to make the vielbein and the
metric dimensionless, $E^m_\mu=m\bar{E}^m_\mu$. This replaces
$\epsilon\rightarrow\bar{\epsilon}=\epsilon m^d~,~\mu\rightarrow\bar{\mu}=\mu m^{d-2}$.
The precise relation between $\mu$ and Newton's constant will be given in
sect. \ref{linearizedfield}. In the
following we will assume that this rescaling has been done and omit the bars on $E^m_\mu,
\mu,$ and $\epsilon$ such that $\mu$ and $\epsilon$ have dimension mass$^{d-2}$ and
mass$^d$, respectively. The remaining two dimensionless parameters $\tau_A$ and $\beta_A$
account for possible deviations from Einstein's gravity. We will find that tight
observational bounds exist only for the parameter $\beta_A$. It is therefore
very interesting that the one loop contribution to $\beta_A$ vanishes \cite{HW1}.

We close this section by noting that the three invariants can also be interpreted
in terms of torsion. Indeed, we may define a different connection
$\tilde{\Gamma}_{\mu\nu}^{\ \ \ \lambda}$ and a new covariant derivative $\tilde{D}_\mu$
such that the vielbein is covariantly conserved
\begin{equation}\label{55QA}
\tilde{D}_\mu E^m_\nu=\partial_\mu E^m_\nu-\tilde{\Gamma}_{\mu\nu}^{\ \ \ \lambda}
E^m_\lambda=0.
\end{equation}
This fixes the connection as\footnote{This connection is often called Weizenb\"ock connection and dsicussed in the context of teleparallel theories\cite{TP}. We stress, however, that the usual teleparallel theories are equivalent reformulations of Einstein's gravity, in contrast to the present work.}
\begin{equation}\label{55QB}
\tilde{\Gamma}_{\mu\nu}^{\ \ \ \lambda}=(\partial_\mu E^m_\nu)E^\lambda_m
\end{equation}
and comparison with eq. (\ref{S10}) identifies \footnote{
Since the l.h.s. of eq. (\ref{55QC}) is a tensor this shows that
$\tilde{\Gamma}_{\mu\nu}{^\lambda}$ indeed transforms as a connection under general
coordinate transformations. Of course, this can also be checked by direct computation
from the analogue of eq. (\ref{4}).} the contorsion
\begin{equation}\label{55QC}
E^\lambda_mD_\mu E^m_\nu=\tilde{\Gamma}_{\mu\nu}^{\ \ \ \lambda}
-\Gamma_{\mu\nu}^{\ \ \ \lambda}
=L_{\mu\nu}^{\ \ \ \lambda}.
\end{equation}
We note that the antisymmetric part of $\tilde{\Gamma}_{\mu\nu}^{\ \ \ \lambda}$ is
the torsion tensor \footnote{In ref. \cite{CWGG} the torsion tensor $T_{\mu\nu\rho}$ is
denoted by $R_{\mu\nu\rho}$}
\begin{equation}\label{55QD}
\tilde{\Gamma}_{\mu\nu}^{\ \ \ \lambda}-\tilde{\Gamma}_{\nu\mu}^{\ \ \ \lambda}
=-2\Omega_{\mu\nu}^{\ \ \ m}E^\lambda_m
=T_{\mu\nu}^{\ \ \ \lambda}.
\end{equation}
For the invariant $I_3$ we observe the identity
\begin{equation}\label{55QE}
D^\mu E^m_\mu=(\tilde{\Gamma}_\mu^{\ \ \mu\lambda}-\Gamma_\mu^{\ \ \mu\lambda})
E^m_\lambda.
\end{equation}
Since eq. (\ref{55QA}) implies the existence of
$d$ covariantly conserved vector fields the connection
$\tilde{\Gamma}$ (\ref{55QB}) is curvature free.

\section{Curvature scalar and field equations}
\label{curvaturescalar}
Of course, the usual curvature scalar $R$ can be constructed from the metric $g_{\mu\nu}$
and the connection $\Gamma$ such that $R[g_{\mu\nu}]$ is invariant.
With $g_{\mu\nu}[E^m_\rho]$ given by eq. (\ref{18AB}) this yields
another invariant involving two derivatives of the vielbein, namely
\begin{equation}\label{S11/1}
\Gamma_{2,R}=\int d^dxE~R\big[g_{\mu\nu}[E^m_\rho]\big].
\end{equation}
The invariants (\ref{S8})(\ref{S11/1}) are not linearly independent,
however:
\begin{equation}\label{S11A}
\Gamma_{2,1}-\Gamma_{2,3}=-2\Gamma_{2,2}+\Gamma_{2,R}~,~
\Gamma_{2,R}=-2I_1.
\end{equation}
This follows by partial integration and use of the commutator identity for two covariant
derivatives
\begin{equation}\label{S11B}
[D_\rho,D_\sigma]E^m_\mu=R_{\rho\sigma\mu}^{\ \ \ \ \nu}E^m_\nu.
\end{equation}

For practical computational purposes it is sometimes convenient to use an alternative
parameterization of $\Gamma$ with
$\Gamma_{2,R},\Gamma_{2,1}$ and $\Gamma_{2,2}$ as independent invariants.
We may rewrite the effective action in the form
\begin{equation}\label{S11C}
\Gamma=\epsilon\Gamma_0-\delta\Gamma_{2,R}+{\zeta}\Gamma_{2,1}+\xi\Gamma_{2,2}
\end{equation}
where
\begin{equation}\label{31FF}
\zeta=\tau_A\mu~,~\xi=(\beta_A-\tau_A)\mu~,~\delta=\frac{1}{2}(1-\tau_A)\mu.
\end{equation}
In analogy to the Einstein-equation we can then write the field equation in absence of
sources in the form
\begin{equation}\label{S11D}
2\delta \left(R_{\mu\nu}-\frac{1}{2}Rg_{\mu\nu}\right)=
\hat{T}_{\mu\nu}-\epsilon g_{\mu\nu}.
\end{equation}
Here the contribution from the invariants $\Gamma_{2,1}$ and $\Gamma_{2,2}$ is
formally written as a part of the energy momentum tensor
\begin{eqnarray}\label{XF1}
\hat{T}_{\mu\nu}&=& \zeta\{4\Omega_{\mu\rho m}\Omega_\nu^{\ \ \rho m}-
2(D_\rho\Omega^\rho_{\ \ \nu m})E^m_\mu-\Omega_{\sigma\rho}^{\ \  m}
\Omega^{\sigma\rho}_{\ \ m}
g_{\mu\nu}\}\nonumber\\
&&+\xi \{ E^\sigma_m\partial_\sigma(D^\rho E^m_\rho)g_{\mu\nu}-\partial_\mu
(D^\rho E^m_\rho)E_{\nu m}\nonumber\\
&&+\frac{1}{2}D^\sigma E^m_\sigma
D^\rho E_{\rho m}g_{\mu\nu}\}.
\end{eqnarray}
Details of the derivation of $\hat{T}_{\mu\nu}$ can be found in the appendix A.
The tensor $\hat{T}_{\mu\nu}$ can be decomposed into a symmetric and antisymmetric part
\begin{equation}\label{33aa}
\hat{T}_{\mu\nu}=\hat{T}^{(s)}_{\mu\nu}+\hat{T}^{(a)}_{\mu\nu}
\end{equation}
and the field equation (\ref{S11D}) implies that the antisymmetric part must vanish
in the absence of sources
\begin{equation}\label{33bb}
\hat{T}^{(a)}_{\mu\nu}=\frac{1}{2}(\hat{T}_{\mu\nu}-\hat{T}_{\nu\mu})=0.
\end{equation}
Due to the Bianchi identity the symmetric part is convariantly conserved
\begin{equation}\label{33cc}
D_\nu\hat{T}^{(s)\mu\nu}=0.
\end{equation}

We note that $\hat{T}_{\mu\nu}$ can contain a piece proportional to the Einstein tensor
$R_{\mu\nu}-\frac{1}{2}Rg_{\mu\nu}$. The definition of the ``gravitational energy
momentum tensor'' $\hat{T}_{\mu\nu}$ is therefore not unique and eq. (\ref{XF1})
should be considered as a formal tool. In particular, for $\tau_A\neq 0$
the coefficient $2\delta$ should not be associated with the Planck mass which is rather
related to $\mu$. If one chooses to collect the gravitational effects beyond Einstein
gravity in a gravitational energy momentum tensor a better definition would collect the
pieces from $I_2,I_3$ instead of $\Gamma_{2,1},\Gamma_{2,2}$. This subtracts from
$\hat{T}_{\mu\nu}$ (\ref{XF1}) a piece $(2\delta -\mu)(R_{\mu\nu}-\frac{1}{2}Rg_{\mu\nu})$.
In presence of matter fluctuations (i.e. from the spinor fields) or
expectation values of composite fields beyond the vielbein the energy momentum
tensor will receive additional contributions,
$\hat{T}_{\mu\nu}\rightarrow\hat{T}_{\mu\nu}+T_{\mu\nu}$.

\section{Linearized gravity}
\label{linearizedgravity}
In the next sections we will study the possible phenomenological
consequences of the generalized gravity (\ref{27NNA}).
We start the investigation of possible observable effects with a discussion
of weak gravity. This will also reveal the ``particle content'' of this theory.
The spectrum of small fluctuations around flat space can be investigated by linearization
(for vanishing cosmological constant $\epsilon =0$). In the linear approximation we write
\begin{eqnarray}\label{LL1}
E^m_\mu&=&\delta^m_\mu+\delta E_\mu^{\ \ m}~,
~E^\mu_m=\delta^\mu_m+\delta E_m^{\ \mu}~,\nonumber\\
\delta E_\mu{^m}&=&\frac{1}{2}k_{\mu\nu}\eta^{\nu m}=\frac{1}{2}(h_{\mu\nu}+a_{\mu\nu})
\eta^{\nu m}\nonumber\\
\delta E_m^{\ \mu}&=&-\frac{1}{2}(h_{\rho\nu}+a_{\rho\nu})\delta^\rho_m\eta^{\nu\mu}
\end{eqnarray}
with symmetric and antisymmetric parts
\begin{equation}\label{LL2}
h_{\mu\nu}=h_{\nu\mu}~,~a_{\mu\nu}=-a_{\nu\mu}
\end{equation}
and
\begin{equation}\label{LL3}
g_{\mu\nu}=\eta_{\mu\nu}+h_{\mu\nu}.
\end{equation}
In this and in the next two sections we raise and lower the indices of
$h_{\mu\nu}$ and $a_{\mu\nu}$ with
$\eta^{\mu\nu}~,~\eta_{\mu\nu}$ such that
\begin{eqnarray}\label{LL4}
\delta E_m^{\ \mu}&=&-\frac{1}{2}\eta_{m\rho}(h^{\rho\mu}+a^{\rho\mu})~,~\nonumber\\
g^{\mu\nu}&=&\eta^{\mu\nu}-h^{\mu\nu}.
\end{eqnarray}
We observe that the antisymmetric fluctuation $a_{\mu\nu}$ does not contribute to the
fluctuation of the metric. In linear order one finds
\begin{eqnarray}\label{LL5}
\Omega_{\mu\nu}^{\ \ \ m}&=&-\frac{1}{4}\eta^{\rho m}
(\partial_\mu h_{\nu\rho}-\partial_\nu h_{\mu\rho}
+\partial_\mu a_{\nu\rho}-\partial_\nu a_{\mu\rho})\\
D^\mu E^m_\mu&=&\frac{1}{2}\eta^{\rho m}
(\partial^\mu a_{\mu\rho}-\partial^\mu h_{\mu\rho}+\partial_\rho h^\mu_\mu)\label{LL6}
\end{eqnarray}
and the invariants in quadratic order therefore read
\begin{eqnarray}\label{LL7}
\Gamma_{2,1}&=&\frac{1}{8}\int d^dx\{\partial^\mu a^{\nu\rho}\partial_\mu a_{\nu\rho}
-\partial^\mu a^{\nu\rho}\partial_\nu a_{\mu\rho}\nonumber\\
&&-2\partial^\mu h^{\nu\rho}\partial_\nu a_{\mu\rho}+\partial^\mu h^{\nu\rho}\partial_\mu
h_{\nu\rho}-\partial^\mu h^{\nu\rho}\partial_\nu h_{\mu\rho}\}
\end{eqnarray}
and
\begin{eqnarray}\label{LL8}
\Gamma_{2,2}&=&\frac{1}{8}\int d^dx(\partial_\mu a^{\mu\rho}-\partial_\mu h^{\mu\rho}
+\partial^\rho h^\mu_\mu)\nonumber\\
&&(\partial^\nu a_{\nu\rho}-\partial^\nu h_{\nu\rho}+\partial_\rho h^\nu_\nu).
\end{eqnarray}

We next decompose $h_{\mu\nu}$ and $a_{\mu\nu}$ in orthogonal irreducible representations
of the Poincare group
\begin{eqnarray}\label{LL9}
h_{\mu\nu}&=&\sum^4_{k=1}h^{(k)}_{\mu\nu}=\sum^4_{k=1}
(P^k)^{\rho\sigma}_{\mu\nu} h_{\rho\sigma}~,\nonumber\\
a_{\mu\nu}&=&\sum^2_{l=1} a^{(l)}_{\mu\nu}
=\sum^2_{l=1}(\bar{P}^l)^{\rho\sigma}_{\mu\nu} a_{\rho\sigma}.
\end{eqnarray}
Here $(\partial^2=\eta^{\rho\sigma}\partial_\rho\partial_\sigma)$
\begin{eqnarray}\label{LL10}
h^{(1)}_{\mu\nu}&=&b_{\mu\nu}~,~h^{(2)}_{\mu\nu}=\frac{1}{d-1}\left(\eta_{\mu\nu}
-\frac{\partial_\mu\partial_\nu}{\partial^2}\right)\sigma~,\nonumber\\
h^{(3)}_{\mu\nu}&=&\frac{\partial_\mu\partial_\nu}{\partial^2}f~,~
h^{(4)}_{\mu\nu}=\partial_\mu v_\nu+\partial_\nu v_\mu~,\nonumber\\
a^{(1)}_{\mu\nu}&=&c_{\mu\nu}~,~a^{(2)}_{\mu\nu}=\partial_\mu
(v_\nu+w_\nu)-\partial_\nu(v_\mu+w_\mu)
\end{eqnarray}
obey the constraints
\begin{eqnarray}\label{LL11}
\partial_\mu b^{\mu\nu}&=&0~,~\eta^{\mu\nu} b_{\mu\nu}=0~,\nonumber\\
\partial_\mu v^\mu&=&0~,~\partial_\mu c^{\mu\nu}=0~,~\partial_\mu w^\mu=0
\end{eqnarray}
such that
\begin{eqnarray}\label{LL12}
h^\mu_\mu=\sigma+f~,~\partial_\mu h^{\mu\nu}=\partial^2 v^\nu+\partial^\nu f,\nonumber\\
\partial_\mu\partial_\nu h^{\mu\nu}=\partial^2 f~,~\partial_\mu a^{\mu\nu}=
\partial^2(v^\nu+w^\nu).
\end{eqnarray}
This yields the effective action (\ref{27NNA}) in quadratic order $(\epsilon=0)$
\begin{eqnarray}\label{LL14}
\Gamma&=&\frac{\mu}{8}\int d^dx\left\{\partial^\mu b^{\nu\rho}\partial_\mu b_{\nu\rho}
-\left(\frac{d-2}{d-1}-\beta_A\right)\partial^\mu\sigma\partial_\mu\sigma\right.\nonumber\\
&&\left.+\tau_A\partial^\mu c^{\nu\rho}\partial_\mu c_{\nu\rho}
+\beta_A\partial^2 w^\mu\partial^2w_\mu \right.\Big\}.
\end{eqnarray}

We observe that $f$ and $v_\mu$ are pure gauge degrees of freedom and do not appear in
$\Gamma$. In addition to the usual metric degrees of freedom $b_{\mu\nu}$ and
$\sigma$ spinor gravity contains the new massless fields $c_{\mu\nu}$ and
$w_\mu$. In presence of a local Lorentz symmetry (as in the usual setting) $c_{\mu\nu}$ and
$w_\mu$ would be the gauge degrees of freedom of the local Lorentz group. Here they rather
have the character of Goldstone degrees of freedom associated to the spontaneous
breaking of the global Lorentz symmetry. In fact, the vielbein of a flat ground state,
$E^m_\mu=\delta^m_\mu$, spontaneously breaks the global rotations acting on the
index $m$. It remains invariant, however, under a combined global Lorentz transformation
and coordinate rotation, the latter acting on the index $\mu$.
We note that for $\beta_A=0$ the only
modification of standard gravity would be the additional antisymmetry field $c_{\mu\nu}$.
This leads to the speculation that within spinor gravity the field $w^\mu$ could
correspond to the gauge degree of freedom of a yet unidentified, perhaps
nonlinear symmetry.
We will see that the coupling $\tau_A$ is not restricted by present observation.

\section{Linearized field equations}
\label{linearizedfield}
For the derivation of the field equations it is useful to write the effective
action (\ref{LL14}) in a form which uses explicitely the projectors. Defining
$k_{\mu\nu}=h_{\mu\nu}+a_{\mu\nu}$, $k^{(k)}_{\mu\nu}=h^{(k)}_{\mu\nu}$ for
$k=1\dots 4~,~k^{(5)}_{\mu\nu}=a^{(1)}_{\mu\nu}~,~k^{(6)}_{\mu\nu}=a^{(2)}_{\mu\nu}$ and
correspondingly $P^5=\bar{P}^1~,~P^6=\bar{P}^2$ one finds
\begin{equation}\label{LL15}
\Gamma=-\frac{\mu}{8}k^{\mu\nu}\partial^2(\sum^6_{k=1}A_kP^k)^{\rho\sigma}_{\mu\nu}
k_{\rho\sigma}
\end{equation}
with
\begin{eqnarray}\label{LL16}
A_1&=&1~,~A_2=-(d-2)+(d-1)\beta_A~,\nonumber\\
A_4&=&0~,~A_5=\tau_A.
\end{eqnarray}
The projectors
\begin{eqnarray}\label{LL17}
P^1&=&\frac{1}{2}(\delta^\rho_\mu\delta^\sigma_\nu
+\delta^\sigma_\mu\delta^\rho_\nu)-\frac{1}{d-1}
\eta_{\mu\nu}\eta^{\rho\sigma}\nonumber\\
&&-\frac{1}{2}\left(\frac{\partial_\mu\partial^\rho}{\partial^2}
\delta^\sigma_\nu
+\frac{\partial_\nu\partial^\rho}{\partial^2}\delta^\sigma_\mu
+\frac{\partial_\mu\partial^\sigma}{\partial^2}\delta^\rho_\nu+
\frac{\partial_\nu\partial^\sigma}{\partial^2}\delta^\rho_\mu\right)\nonumber\\
&&+\frac{1}{d-1}\left(\frac{\partial_\mu\partial_\nu}{\partial^2}\eta^{\rho\sigma}
+\frac{\partial^\rho\partial^\sigma}{\partial^2}\eta_{\mu\nu}\right)\nonumber\\
&&+\frac{d-2}{d-1}\frac{\partial_\mu\partial_\nu\partial^\rho\partial^\sigma}{\partial^4},
\nonumber\\
P^2&=&\frac{1}{d-1}\left(\eta_{\mu\nu}-\frac{\partial_\mu\partial_\nu}{\partial^2}\right)
\left(\eta^{\rho\sigma}-\frac{\partial^\rho\partial^\sigma}{\partial^2}\right),\nonumber\\
P^3&=&\frac{\partial_\mu\partial_\nu\partial^\rho\partial^\sigma}{\partial^4}~,~\nonumber\\
P^4&=&\frac{1}{2}\left(\frac{\partial_\mu\partial^\rho}{\partial^2}\delta^\sigma_\nu
+\frac{\partial_\nu\partial^\rho}{\partial^2}\delta^\sigma_\mu
+\frac{\partial_\mu\partial^\sigma}{\partial^2}\delta^\rho_\nu
+\frac{\partial_\nu\partial^\sigma}{\partial^2}\delta^\rho_\mu\right) \nonumber\\
&&-2\frac{\partial_\mu\partial_\nu\partial^\rho\partial^\sigma}{\partial^4},\nonumber\\
P^5&=&\frac{1}{2}
\left(\frac{\partial_\mu\partial^\rho}{\partial^2}\delta^\sigma_\nu-
\frac{\partial_\nu\partial^\rho}{\partial^2}\delta^\sigma_\mu
-\frac{\partial_\mu\partial^\sigma}{\partial^2}\delta^\rho_\nu
+\frac{\partial_\nu\partial^\sigma}{\partial^2}\delta^\rho_\mu\right),\nonumber\\
P^6&=&\frac{1}{2}\left(\delta^\rho_\mu\delta^\sigma_\nu
-\delta^\sigma_\mu\delta^\rho_\nu\right)-P_5
\end{eqnarray}
obey $(P^k)^2=P^k$ and are orthogonal $P^jP^k=0$ for $j\neq k$.

In order to derive the linear field equations in presence of sources - e.g. matter
concentrations - we add to $\Gamma$ a term involving the symmetric energy-momentum tensor
$T_{\mu\nu}=T_{\nu\mu}$ of matter and radiation
\begin{eqnarray}\label{LL18}
\Gamma_M&=&-\frac{1}{2}\int d^dxh^{\mu\nu}T_{\mu\nu}
=-\frac{1}{2}\int d^dxh_{\mu\nu}T^{\mu\nu}\\
&=&-\frac{1}{2}\int d^dx\left\{b_{\mu\nu}\left(T^{\mu\nu}
-\frac{1}{d-1}T^\rho_\rho\eta^{\mu\nu}\right.\right.\nonumber\\
&&\left.\left.+\frac{1}{d-1}
\frac{\partial_\mu\partial_\nu}{\partial^2}T^\rho_\rho\right)
+\frac{1}{d-1}\sigma T^\rho_\rho\right\}.\nonumber
\end{eqnarray}
Here we have used the linear energy momentum conservation $\partial_\mu T^{\mu\nu}=0$ for
the last line. We will motivate in the next section the omission of the antisymmetric part
of $T_{\mu\nu}$ in more detail.

The field equations follow from the variation of
$\Gamma+\Gamma_M$, eqs. (\ref{LL16}), (\ref{LL18})
\begin{equation}\label{LL19}
-\partial^2\sum_kA_k(P^k)^{\rho\sigma}_{\mu\nu}k_{\rho\sigma}
=\frac{2}{\mu}T_{\mu\nu}.
\end{equation}
It can be projected on the irreducible representations
\begin{equation}\label{LL20}
-\partial^2A_kk^{(k)}_{\mu\nu}=\frac{2}{\mu}(P_k)^{\rho\sigma}_{\mu\nu}T_{\rho\sigma}
\end{equation}
where we note $(P_k)^{\rho\sigma}_{\mu\nu}T_{\rho\sigma}=0$ for
$k=3,4,5,6$ due to the symmetry of $T_{\mu\nu}$ and $\partial_\mu T^{\mu\nu}=0$.
This yields
\begin{eqnarray}\label{LL21}
-\partial^2h^{(1)}_{\mu\nu}&=&\frac{2}{A_1\mu}\left(T_{\mu\nu}-\frac{1}{d-1}
T^\rho_\rho\eta_{\mu\nu}+\frac{1}{d-1}\frac{\partial_\mu\partial_\nu}{\partial^2}
T^\rho_\rho\right)\nonumber\\
-\partial^2h^{(2)}_{\mu\nu}&=&\frac{2}{A_2(d-1)\mu}\left(\eta_{\mu\nu}
-\frac{\partial_\mu\partial_\nu}{\partial^2}\right)T^\rho_\rho
\end{eqnarray}
in accordance \footnote{In this respect it is crucial that $b_{\mu\nu}$
multiplies the correctly projected $(P^1 T)^{\mu\nu}$ in eq. (\ref{LL18}).}
with a variation of eqs. (\ref{LL14}) (\ref{LL18}) with respect to $b_{\mu\nu}$ and
$\sigma$. The remaining field equations can be written as
\begin{equation}\label{LL22}
\tau_A\partial^2c_{\mu\nu}=0~,~\beta_A\partial^2 w_\mu=0
\end{equation}
since nonvanishing sources for $c_{\mu\nu}$ and $w_\mu$ can only arise from the antisymmetric part
of a generalized energy momentum tensor.

We can now compute the Newtonian limit by inserting $T_{\mu\nu}=\rho\delta_{\mu0}
\delta_{\nu0}$ and considering a time independent metric with Newtonian potential
\begin{equation}\label{LL23}
\phi=-\frac{1}{2}h_{00}=-\frac{1}{2}(h^{(1)}_{00}+h^{(2)}_{00}).
\end{equation}
For $d=4$ one has $(\Delta=\partial^i\partial_i)$
\begin{equation}\label{LL24}
-\Delta h^{(1)}_{00}=\frac{4\rho}{3\mu}~,~-\Delta h^{(2)}_{00}=-\frac{\rho}{3\mu}
\left(1-\frac{3}{2}\beta_A\right)^{-1}
\end{equation}
or
\begin{equation}\label{LL25}
\Delta\phi=\frac{\rho}{2\mu}\frac{1-2\beta_A}{1-\frac{3}{2}\beta_A}=4\pi G_N\rho=
\frac{\rho}{2\bar{M}^2}.
\end{equation}
This fixes Newton's constant $G_N$ or the reduced Planck mass $\bar{M}^2=M^2_p/8\pi$ as
\begin{equation}\label{LL26}
\bar{M}^2=\frac{1-\frac{3}{2}\beta_A}{1-2\beta_A}\mu.
\end{equation}
We will see below that $\beta_A$ has to be small such that $\bar{M}^2$ is essentially
given by $\mu$. However, within Newtonian gravity the couplings $\tau_A$ and $\beta_A$
are not constrained.

The linear approximation governs the emission, propagation and detection of gravitational
waves. Those are described by $h^{(1)}_{\mu\nu}=b_{\mu\nu}$. For example, the emission of
gravitational waves from pulsars is the same as in Einstein gravity. However, the effective
reduced Planck mass extracted from the gravitational radiation of pulsars is given by
$\bar{M}^2_{pulsar}=\mu$ (since $A_1=1)$. In view of the tight limit for $|\beta_A|$ derived
in sect. \ref{modificationofthe} the difference between
the gravitational constant measured in Newtonian
gravity (\ref{LL26}) and the one relevant for pulsars seems to be too small in order to be
observable.

Actually, if $\Gamma$ is given by the one loop approximation
one finds that $\beta_A$ vanishes. Indeed, we may compute
$\Omega_{\mu\nu\rho}$ in the linearized approximation
\begin{eqnarray}\label{88Ya}
\Omega_{\mu\nu\rho}&=&-\frac{1}{4}\big\{\partial_\mu h_{\nu\rho}
+\partial_\mu a_{\nu\rho}-(\mu\leftrightarrow\nu)\big\}\nonumber\\
&=&-\frac{1}{4}\big\{\partial_\mu(b_{\nu\rho}+c_{\nu\rho})
-\partial_\nu(b_{\mu\rho}+ c_{\mu\rho})\nonumber\\
&&+\frac{1}{d-1}(\eta_{\nu\rho}\partial_\mu\sigma
-\eta_{\mu\rho}\partial_\nu\sigma)\nonumber\\
&&-\partial_\rho(\partial_\mu w_\nu
-\partial_\nu w_\mu)\big\}.
\end{eqnarray}
The totally antisymmetric part therefore only involves $c_{\nu\rho}$
\begin{equation}\label{88YB}
\Omega_{[\mu\nu\rho]}=-\frac{1}{2}\partial_{[\mu}c_{\nu\rho]}.
\end{equation}
We will see in sect. \ref{generalizeddirac} that
the modified Dirac operator in our generalized gravity
can be written in the form
${\cal D}={\cal D}_E+\frac{1}{4}\Omega_{[\mu\nu\rho]}\gamma^{\mu\nu\rho}_{(3)}$
(\ref{PB22}). Here ${\cal D}_E$ can only depend on $b_{\nu\rho}$ and $\sigma$ as
a consequence of local Lorentz invariance. We conclude that ${\cal D}$ does not depend on
$w_\mu$. In consequence, the one loop expression $Tr\ln{\cal D}$ cannot lead
to a term $\sim \partial^2 w^\mu\partial^2 w_\mu$ in quadratic order.
Comparison with eq. (\ref{LL14}) implies $\beta_A=0$.

\section{Anomalous spin interactions and post-Newtonian gravity}
\label{anomalousspin}
In this section we discuss the anomalous couplings of gravitational degrees of freedom
to the spin of fermions. This is related to possible anomalous spin interactions and
the issue of post-Newtonian gravity beyond the linear approximation. For our purpose
we need information about the coupling of the vielbein $E^m_\mu$ to spinor fields $\psi$.
From symmetry arguments \cite{SG2} one expects that the missing spin connection reflects
itself in a modification of the covariant spinor kinetic term
\begin{equation}\label{AS1}
{\cal L}_\psi=i\bar{\psi}\gamma^\mu D^{(E)}_\mu\psi+
i\sigma_A\Omega_{[\mu\nu\rho]}\bar{\psi}\gamma^{\mu\nu\rho}_{(3)}\psi.
\end{equation}
Here the covariant derivative $D^{(E)}_\mu$ is constructed as usual and involves
the standard spin connection constructed from $E^m_\mu$ and its derivatives
(see sect. \ref{generalizeddirac} for details). The anomalous term proportional
to the coupling $\sigma_A$ reflects the violation of local Lorentz symmetry
in the spinor coupling. In the classical approximation to spinor gravity one has
$\sigma_A=1/4$.

As mentioned already, the first term in eq. (\ref{AS1}) only couples to
$h_{\mu\nu}$. It gives the standard contribution of fermionic particles to the
symmetric energy momentum tensor $T^{\mu\nu}_{(g)}$ (\ref{18AD}). In contrast,
the anomalous second part yields in the linear approximation
\begin{equation}\label{AS2}
{\cal L}_A=-\frac{i}{2}\sigma_A\partial_{[\mu}c_{\nu\rho]}\bar{\psi}
\gamma^{\mu\nu\rho}_{(3)}\psi.
\end{equation}
Using partial integration one finds \cite{HW1} that the coupling of
$c_{\nu\rho}$ to the spinors
\begin{eqnarray}
{\cal L}_A&\sim& c_{\nu\rho}\partial_\mu(\bar{\psi}\gamma^{[\mu}\gamma^\nu\gamma^{\rho]}\psi)
\nonumber\\
&\sim&c_{\nu\rho}\epsilon^{\nu\rho\mu\sigma}\partial_\mu S_\sigma
\end{eqnarray}
involves the density of the spin vector $S_\sigma$. (We recall that there is no
coupling to $a^{(2)}_{\nu\rho}$.) In consequence, the exchange of $c_{\nu\rho}$
induces a dipole-dipole interaction between fermions with infinite range.
However, its strength is only gravitational $\sim \mu^{-1}\sim \bar{M}^{-2}$ and
therefore suppressed as compared to the magnetic dipole interaction by a factor
$\sim(m_e/\bar{M})^2)\sim 10^{-44}$ - many orders of magnitude too small to be observable
\cite{SO}.

We conclude that we can safely neglect the anomalous spinor coupling and
concentrate on the
symmetric energy momentum tensor  $T^{\mu\nu}_{(g)}$. Indeed, as compared
to Newtonian gravity the macroscopic spin forces are doubly suppressed. First, the
lack of spin coherence of macroscopic bodies leads to suppression factors
$S_{tot}\cdot m/M_{tot}$ for each body involved. Here $S_{tot}$ is the total
spin (in units of $\hbar$) and $M_{tot}$ the total mass of the body
composed of particles with (average) mass $m$. Second, a dipole-dipole force decays very
fast $\sim r^{-3}$.

We will next see that the particular form of the coupling of $c_{\nu\rho}$ to the
spinor field also has important consequences for post-Newtonian gravity. All effects from
a violation of the local Lorentz symmetry by the invariant $\sim\tau_A$ are severely suppressed.

The expansion in weak gravitational fields can be extended beyond linear order. A
general framework for the effects beyond Newtonian gravity is given by the
``Parameterized Post-Newtonian Formalism'' (PPN) \cite{Will}. In principle, one should
perform a systematic computation of all PPN parameters for the field equations following
from the effective gravitational action (\ref{AA2}) without local Lorentz-invariance. For
weak gravitational fields the higher order corrections are computed
iteratively: one uses the results of the linear approximation
(cf. the preceeding two sections) in order to derive the field equations for the
derivations of $E^m_\mu$ from the linear approximation. The PPN-formalism needs
at most the next-to-linear terms for some of the vielbein components. Deviations from
Einstein's gravity therefore involve the cubic couplings arising from the invariants
$\sim \tau_A$ and $\beta_A$.

In the remainder of this section we show that for $\beta_A=0$ the invariant $\tau_A$
gives precisely the same PPN-results as Einstein's gravity. The coupling
$\tau_A$ is therefore not constrained by any one of the tests of general relativity
for weak fields (i.e. up to the order used for the PPN-formalism). This
demonstrates that local Lorentz symmetry is actually very poorly tested - an
additional invariant violating local Lorentz symmetry can be added and remains
essentially undetectable by present means! We will not pursue a systematic
PPN-discussion of the influence of the coupling $\beta_A$ since we believe that the
strongest constraints arise for isotropic gravitational fields which will be discussed in
complete nonlinear order in the next sections.

For $\beta_A=0$ there is no disctinction between Einstein gravity and our generalized
gravity in the linear approximation provided $T^{\mu\nu}=T^{\mu\nu}_{(g)}$.
(We have shown above that the neglection of the antisymmetric part of $T^{\mu\nu}$ is
indeed a very good approximation.) In particular, $h_{\mu\nu}$ takes in the linear
approximation the same values as for Einstein's gravity and $a_{\mu\nu}=0$. in principle,
deviations on the PPN level could arise if the invariant $\sim\tau_A$ produces cubic terms
$\sim h^3$ or $h^2a$. Then the field equations for the corrections in next-to-linear order 
would have additional source terms $\sim h^2_l$ where $h_l$ is the metric in linear order.
(Recall $a_l=0$.) We will show that such cubic terms are not present at the PPN level.

By straightforward algebraic manipulations one establishes that the invariant $\sim \tau_A$
only involves the totally antisymmetric part of $\Omega_{\mu\nu\rho}$
\begin{equation}\label{input1}
I_2=\frac{3}{2}\int d^dxE\Omega^{[\mu\nu\rho]}\Omega_{[\mu\nu\rho]}
\end{equation}
Expanding $\Omega_{[\mu\nu\rho]}$ up to terms quadratic in $h$ and $a$ one finds
$(\partial_{[\mu}a_{\nu\rho]}=\partial_{[\mu}c_{\nu\rho]})$
\begin{equation}\label{input2}
\Omega_{[\mu\nu\rho]}=-\frac{1}{2}\partial_{[\mu}a_{\nu\rho]}-G_{[\mu\nu\rho]}+\cdots
\end{equation}
with
\begin{equation}\label{input3}
G_{\mu\nu\rho}=\frac{1}{8}(\partial_\mu h_{\nu\sigma})h_{\rho\tau}\eta^{\sigma\tau}
\end{equation}
The dots denote terms $\sim ah$ and $a^2$ which are not relevant for our purpose.
The crucial point is that $h$ appears only in quadratic order in
$\Omega_{[\mu\nu\rho]}$ and $I_2$ therefore contains no term $\sim h^3$. On the PPN level
possible modification can therefore only arise from
\begin{equation}\label{input4}
I^{(3)}_2=\int d^dx G^{[\mu\nu\rho]}\partial_{[\mu}c_{\nu\rho]}
\end{equation}
This term results in a gravitational source term for $c_{\nu\rho}$ which is $\sim h^2_l$.

On the PPN-level the source $\sim \partial_\mu G^{[\mu\nu\rho]}$ vanishes. Since the source
is already $\sim h^2$ we only need to take into account the Newtonian part $(h_l)_{00}$.
In this case one infers $G_{[\mu\nu\rho]}=0$ since an antisymmetrization over two equal
indices $\nu=\rho=0$ is involved. This concludes our argument. Can one conceive future experiments
that could detect the field $c_{\nu\rho}$ as a manifestation of the violation of local
Lorentz symmetry for the case $\beta_A=0$? The answer tends to be negative: it is simply
very hard to produce a macroscopic $c_{\nu\rho}$-field with observable strength. And even
if one would succeed the measurement would require a probe with coherent spin.

\section{General isotropic static solution}
\label{generalisotropic}
The comparison with Einstein gravity should, of course, not be restricted to the linear
approximation. The two prominent examples where nonlinear effects play a role are the
Schwarzschild solution and cosmology. They will be discussed in sects. \ref{generalisotropic}
-
\ref{homogeneousisotropic}. In
this section we discuss the Schwarzschild solution for the generalized
gravity corresponding to the effective action (\ref{27NNA}) or (\ref{S11C}). For this
purpose we describe the most general static solution of the nonlinear gravitational
field equations under the assumption of isotropy. Obviously, this
goes beyond Newtonian gravity and linearized gravity.
In sect. \ref{modificationofthe} we concentrate
on $d=4$ and compare our general solution with the Schwarzschild solution in Einstein
gravity. For $\beta_A=0$ we find that the standard Schwarzschild solution is also a
solution to the nonlinear field equation of our generalized gravity. The parameter $\tau_A$
therefore remains unconstrained. On the other hand, for $\beta_A\neq 0$ we find a
difference already in the lowest order of the post-Newtonian expansion. Recent
precision observations put a severe bound on the parameter $\beta_A$.

A rotation acts on the vielbein $E^m_\mu$ as a coordinate rotation leaving
$r^2=\sum^{d-1}_{i=1}x^2_i$ invariant, accompanied by a simultaneous suitable global
Lorentz rotation acting on the index $m$. The most general rotation invariant vielbein
takes the form \footnote{There should be no confusion between $f(r)$ and the gauge degree of
freedom $f$ in sect. \ref{linearizedgravity}.}
\begin{eqnarray}\label{R1}
E^0_0=f(r)&,&E^j_0=h(r)x_j~,\nonumber\\
E^0_i=g(r)x_i&,&E^j_i=c(r)\delta_{ij}+k(r)x_ix_j.
\end{eqnarray}
We can rescale $r=D(r^\prime)r^\prime$ in order to fix $c(r)=1$. Similarly, a
radius dependent rescaling of the clocks $dt=dt^\prime+rF(r)dr$ leaves $d\xi^m=
E^m_\mu dx^\mu=E^{\prime m}_\mu dx^{\prime\mu}$ invariant
\begin{eqnarray}\label{R2}
d\xi^0&=&f(r)dt+g(r)rdr\nonumber\\
&=&f(r)dt^\prime+\Big(f(r)F(r)+g(r)\Big)rdr\nonumber\\
&=&f(r)dt^\prime+g^\prime(r)rdr~,\nonumber\\
d\xi^i&=&h(r)x_idt+dx_i+k(r)rx_idr\nonumber\\
&=&h(r)x_idt^\prime+dx_i+\big(k(r)+h(r)F(r)\big)rx_idr\nonumber\\
&=&h(r)x_idt^\prime+dx_i+k^\prime(r)rx_idr.
\end{eqnarray}
With $g^\prime=g+fF,k^\prime=k+hF$ we can use this freedom in order to fix $k$ as a
function of $f,g$ and $h$ such that \footnote{A suitable function $F$ exists provided
$r^2h^2\neq f^2$}
\begin{equation}\label{R3}
gf=h(1+r^2k).
\end{equation}
With this coordinate choice we remain with three free functions $f(r),g(r)$ and
$h(r)$.

From eqs. (\ref{R1}) with (\ref{R3}) and $c=1$ we can compute the metric
\begin{eqnarray}\label{R4}
g_{00}&=&-B(r)~,~g_{0i}=0~,~\nonumber\\
g_{ij}&=&\delta_{ij}+\frac{A(r)-1}{r^2}x_ix_j
\end{eqnarray}
which corresponds for $d=4$ to the line element of the Schwarzschild metric in standard
(polar) coordinates,
\begin{equation}\label{R4a}
ds^2=-B(r)dt^2+A(r)dr^2+r^2(d\vartheta^2+sin^2\vartheta d\varphi^2).
\end{equation}
Here we have used eq. (\ref{R3}) in order to obtain $g_{0i}=0$. The functions
$A(r),B(r)$ are related to $f(r),g(r),h(r)$ by
\begin{equation}\label{R5}
B=f^2-r^2h^2~,~A=1-r^2g^2+2r^2k+r^4k^2=\frac{g^2}{h^2}B.
\end{equation}
We choose as the three independent functions $A(r),B(r)$ and $h(r)$ with
$f=\sqrt{B+r^2h^2},g=\sqrt{A/B}\cdot h$ and $k(r)=r^{-2}(\sqrt{A+r^2h^2A/B}-1)$.
We emphasize that we have here one more free function $h(r)$ in addition to the
two functions $A(r),B(r)$ characterizing the metric. In case of {\em local}
Lorentz symmetry $h(r)$ would correspond to a gauge degree of freedom - here it is not.

Inserting the vielbein
\begin{eqnarray}\label{RZ1}
E^0_0&=&\sqrt{B+r^2h^2}~,~E^j_0=hx_j~,~\nonumber\\
E^0_i&=&h\sqrt{A/B}~x_i~,\nonumber\\
E_{ij}&=&\delta_{ij}+\frac{x_ix_j}{r^2}(\sqrt{A+r^2h^2A/B}-1)
\end{eqnarray}
we compute in appendix A the invariants
\begin{eqnarray}\label{RZ2}
Y_{2,1}&=&\Omega_{\mu\nu}^{\ \ m}\Omega^{\mu\nu}_{\ \ m}\nonumber\\
&=&\frac{d-2}{2r^2}
\left(1+\frac{1}{A}-2
\sqrt{\frac{B+r^2h^2}{AB}}\right)\nonumber\\
&+&\frac{1}{8AB(B+r^2h^2)}
\Big[B^{\prime 2}+4rB^{\prime}h(h+rh^\prime)\nonumber\\
&-&4B(h+rh^\prime)^2\Big]
\end{eqnarray}
and
\begin{eqnarray}\label{R14}
Y_{2,2}&=&\frac{1}{2}(D^\mu E^m_\mu)(D^\nu E_{\nu m})\nonumber\\
&=&\frac{(d-2)^2}{2r^2}
\left(\sqrt{\frac{B+r^2h^2}{AB}}-1\right)^2\nonumber\\
&&+\frac{d-2}{2rAB}\left(1-\sqrt{\frac{AB}{B+r^2h^2}}\right)
\big[B^\prime+2rh(h+rh^\prime)\big]\nonumber\\
&&+\frac{1}{8AB(B+r^2h^2)}\big[B^\prime+2rh(h+rh^\prime)\big]^2\nonumber\\
&&-\frac{1}{2AB}\big[(d-1)h+rh^\prime\big]^2.
\end{eqnarray}
With $E=\sqrt{AB}$ the effective action (\ref{S11C}) becomes
\begin{equation}\label{R15}
\Gamma=\int d^dx\sqrt{AB}\{\epsilon-\delta R[A,B]+\zeta Y_{2,1}+\xi Y_{2,2}\},
\end{equation}
where (cf. appendix A)
\begin{eqnarray}\label{76AA}
R[A,B]&=&-\frac{B^{\prime\prime}}{AB}+\frac{B^\prime}{2AB}\left(\frac{A^\prime}{A}
+\frac{B^\prime}{B}\right)\nonumber\\
&&+\frac{d-2}{rA}
\left(\frac{A^\prime}{A}-\frac{B^\prime}{B}\right)\nonumber\\
&&+\frac{(d-3)(d-2)}{r^2}
\left(1-\frac{1}{A}\right).
\end{eqnarray}

Since our ansatz covers the most general rotation invariant static vielbein (up to
coordinate transformations) the field equations for this symmetric situation
can be obtained by varying $\Gamma$ with respect to $A,B$ and $h$.
For small $h$ we observe that $Y_{2,1}$ and $Y_{2,2}$ are quadratic in $h$ (or
derivatives of $h$). The field equation $\delta\Gamma/\delta h=0$ admits therefore
always solutions with $h=0$. In this situation the remaining field equations for $A$ and
$B$ can be obtained by inserting $h=0$ into $Y_{2,1}$ and $Y_{2,2}$
\begin{eqnarray}\label{R16}
Y_{2,1}(h=0)&=&\frac{d-2}{2r^2}\left(1-\sqrt{\frac{1}{A}}\right)^2
+\frac{1}{8A}\left(\frac{B^\prime}{B}\right)^2,\nonumber\\
Y_{2,2}(h=0)&=&\frac{1}{2}\left[\frac{d-2}{r}\left(1-\sqrt{\frac{1}{A}}\right)
-\frac{1}{2\sqrt{A}}\frac{B^\prime}{B}\right]^2.
\end{eqnarray}

\section{Modification of the Schwarzschild solution}
\label{modificationofthe}
In the following we concentrate on $d=4$ where the resulting effective action reads
after partial integration $(\epsilon=0)$
\begin{eqnarray}\label{R17}
\Gamma[A,B]&=&8\pi\delta\int dtdr\sqrt{AB}\Bigg\{\frac{1}{A}-1-\frac{rA^\prime}{A^2}
\nonumber\\
&&+\tilde{\zeta}\left[\left(1-\frac{1}{\sqrt{A}}\right)^2
+\frac{r^2}{8A}
\left(\frac{B^\prime}{B}\right)^2\right]\nonumber\\
&&+2\tilde{\xi}\left(1-\frac{1}{\sqrt{A}}-\frac{r}{4\sqrt{A}}
\frac{B^\prime}{B}\right)^2\Bigg\}.
\end{eqnarray}
The contributions
$\sim\tilde{\zeta}=\zeta/(2\delta),\tilde{\xi}=\xi/(2\delta)$ do not vanish. For
$\tilde{\zeta}>0,\tilde{\xi}\geq 0$ they give a strictly positive contribution
to $\Gamma$ whenever $A\neq1,B^\prime\neq0$, i.e. for all geometries which differ
from a flat space-time. Standard gravity is recovered for \footnote{This condition
is sufficient but not necessary.} $\tilde{\zeta}=\tilde{\xi}=0$
and leads to the well known Schwarzschild solution $B=A^{-1}=1-r_s/r$ with
Schwarzschild radius $r_s=2mG_N$ related to the total mass $m$ of the object.

The spherically symmetric static field equations obtain by variation of
$\Gamma[A,B]$ with respect to $A$ and $B$. Equivalently, we may express
\footnote{There should be no confusion of $\beta(\rho)$ and the
coupling constant $\beta_A$ in the effective action (\ref{27NNA}) .} $\Gamma$ in terms
of two new functions
$\alpha(\rho),\beta(\rho)$ and a rescaled radial coordinate $\rho$
\begin{equation}\label{R18}
\alpha=A^{-1/2}~,~\beta=\ln B~,~\rho=\ln(r/r_s)
\end{equation}
as
\begin{eqnarray}\label{R19}
\Gamma[\alpha,\beta]&=&8\pi\delta r_s\int dtd\rho e^\rho e^{\frac{\beta}{2}}\left\{
\alpha-\frac{1}{\alpha}+2\dot{\alpha}\right.\nonumber\\
&&+\tilde{\zeta}\left[\frac{(1-\alpha)^2}{\alpha}+\frac{\alpha}{8}\dot{\beta}^2\right]
\nonumber\\
&&\left.+\frac{2\tilde{\xi}}{\alpha}
\left(1-\alpha-\frac{\alpha}{4}\dot{\beta}\right)^2 \right\}.
\end{eqnarray}
Here a dot denotes a derivative with respect to $\rho$. The field equation from the variation
with respect to $\beta$ reads
\begin{eqnarray}\label{R20}
\alpha^2&-&1+2\dot{\alpha}\alpha=\nonumber\\
&-&\tilde{\zeta}\left\{(1-\alpha)^2-\frac{1}{8}
\alpha^2\dot{\beta}^2-\frac{1}{2}\alpha^2(\dot{\beta}+\ddot{\beta})-
\frac{1}{2}\alpha\dot{\alpha}\dot{\beta}\right\}\nonumber\\
&+&\tilde{\xi}\left\{2\alpha\dot{\alpha}+\frac{1}{2}\alpha(\dot{\alpha}
\dot{\beta}+\alpha\ddot{\beta})\right.\nonumber\\
&&\hspace{0.2cm}-\left.\left(2+\frac{1}{2}\alpha\dot{\beta}\right)
\Big(1-\alpha-\frac{\alpha\dot{\beta}}{4}\Big)\right\}
\end{eqnarray}
and from the variation with respect to $\alpha$ we obtain
\begin{eqnarray}\label{R21}
1-\frac{1}{\alpha^2}+\dot{\beta}&=&\nonumber\\
\tilde{\zeta}
\left(1-\frac{1}{\alpha^2}+\frac{\dot{\beta}^2}{8}\right)
&+&2\tilde{\xi}\left[\left(1+\frac{\dot{\beta}}{4}\right)^2-\frac{1}{\alpha^2}\right].
\end{eqnarray}

We may first recover the usual Schwarzschild solution for
$\tilde{\zeta}=\tilde{\xi}=0$ where
\begin{equation}\label{R22}
\frac{\partial\alpha^2}{\partial\rho}=1-\alpha^2~,~
\frac{\partial\beta}{\partial\rho}=\frac{1}{\alpha^2}-1
\end{equation}
implies the solution \footnote{The two integration constants of the general
solution of eq. (\ref{R22}) are an additive constant in $\rho$ which is absorbed in
$r_s$ and a multiplicative constant in $B$ which can be set to one by appropriate
time rescaling.}
\begin{equation}\label{R23}
\alpha^2=e^\beta=1-e^{-\rho}~,~\frac{1}{A}=B=1-\frac{r_s}{r}.
\end{equation}
For general $\tilde{\zeta},\tilde{\xi}$ we make for large $\rho$ the ansatz
\begin{equation}\label{R24}
\alpha^2=1-\gamma e^{-\sigma\rho}~,~\ln\beta=1-e^{-\sigma\rho}.
\end{equation}
Dropping terms $\sim e^{-2\sigma\rho}$ or smaller eqs. (\ref{R20})(\ref{R21}) result in
\begin{eqnarray}\label{R25}
\gamma&-&\gamma\sigma+\frac{1}{2}\tilde{\zeta}(\sigma-\sigma^2)+
\tilde{\xi}\left(\gamma\sigma-\gamma+\frac{\sigma}{2}-\frac{\sigma^2}{2}\right)=0,
\nonumber\\
\gamma&-&\sigma-\tilde{\zeta}\gamma-\tilde{\xi}(2\gamma-\sigma)=0
\end{eqnarray}
with solution
\begin{eqnarray}\label{R26}
\sigma&=&1~,\nonumber\\
\gamma&=&\frac{1-\tilde{\xi}}{1-\tilde{\zeta}-2\tilde{\xi}}
=\left(1-\frac{\tilde{\zeta}+\tilde{\xi}}{1-\tilde{\xi}}\right)^{-1}
=\frac{1-\beta_A}{1-2\beta_A}.
\end{eqnarray}
One finds for $r \gg r_s$ a behavior similar as for a Jordan-Brans-Dicke
\cite{JBD} theory
\begin{equation}\label{R27}
B=1-\frac{r_s}{r}~,~A^{-1}=1-\gamma\frac{r_s}{r}.
\end{equation}
As expected, Newtonian gravity (encoded in $B(r)$) remains unaffected. On the
other hand, post-Newtonian gravity is modified. Strong observational bounds from the
solar system imply that $\gamma$ must be very close to one \cite{Ber}
\begin{equation}\label{R28}
\gamma-1\approx\beta_A\approx\tilde{\zeta}+\tilde{\xi}=
(2.1\pm2.3)10^{-5}.
\end{equation}
This seems to be the most stringent bound on the parameter $\beta_A$.

For $\beta_A=0$, as suggested by the one loop approximation, one has  $\gamma=1$ and there
is no correction to lowest order post-Newtonian gravity. This extends to the full
Schwarzschild solution. Indeed, for $\tilde{\xi}=-\tilde{\zeta}$ the field equations
(\ref{R20}) (\ref{R21}) reduce to the standard field equations (\ref{R22}) for
arbitrary values of $\tau_A$. For $\beta_A=0$ the spherically symmetric static solution does not
distinguish between our version of generalized gravity and Einstein gravity!

Finally, we observe that for $\beta_A\neq0$ the
full solution can be found numerically whereby the initial value problem
has one more free integration constant since
eq. (\ref{R20}) also involves $\ddot{\beta_A}(\rho)$. It would be interesting to investigate
if this has an effect on the singularity. Furthermore, the
solutions with $h=0$ are not the only candidates for the description of the
gravitational effects of massive bodies. One may explore a behavior for $r\gg r_s$ where
\begin{equation}\label{R29}
h(r)=\eta r_s^{1/2}r^{-3/2}.
\end{equation}
We postpone the analysis of such modified solutions to a future investigation.

\section{Homogeneous isotropic cosmology}
\label{homogeneousisotropic}
In this section we investigate cosmologies with a homogeneous and isotropic metric and
a flat spatial hypersurface, i.e. a situation where the Killing vectors of the
(three) spatial translations commute. Again, this tests our generalized gravity beyond the
linear approximation. The most general vielbein consistent with
these symmetries can be brought to the form
\begin{equation}\label{C1}
E^0_0=1~,~E^i_0=0~,~E^0_i=0~,~E^j_i=a(t)\delta^j_i.
\end{equation}
Here $a(t)$ is the usual scale factor which is related to the Hubble parameter by
$H(t)=\dot{a}(t)/a(t)$. We concentrate on three space and one time dimensions,
$d=4$. The
nonvanishing components of the metric and affine connection read
\begin{equation}\label{C2}
g_{00}=-1~,~g_{ij}=a^2(t)\delta_{ij}
\end{equation}
and
\begin{equation}\label{C3}
\Gamma_{ij}^{\ \ \ 0}=Hg_{ij}~,~\Gamma_{0i}^{\ \ \ j}=H\delta^j_i.
\end{equation}
They result in a curvature scalar
\begin{equation}\label{C4}
R=12H^2+6\dot{H}.
\end{equation}
(In this section dots denote time derivatives.) The only independent nonvanishing
components of $\Omega_{\mu\nu}^{\ \ \ m}$ are
\begin{equation}\label{C5}
\Omega_{i0}^{\ \ \ j}=-\Omega_{0i}^{\ \ \ j}=\frac{1}{2}\dot{a}\delta^j_i
\end{equation}
and therefore
\begin{equation}\label{C6}
Y_{2,1}=\Omega_{\mu\nu}^{\ \ \ m}\Omega^{\mu\nu}_{\ \ \ m}=-\frac{3}{2}H^2.
\end{equation}
With
\begin{equation}\label{C7}
D^\mu E^0_\mu=-3H~,~D^\mu E^i_\mu=0
\end{equation}
the second invariant becomes
\begin{equation}\label{C8}
Y_{2,2}=\frac{1}{2}D^\mu E^m_\mu D^\nu E_{\nu m}=-\frac{9}{2}H^2.
\end{equation}

Since we consider the most general vielbein consistent with the symmetries we can
again derive the relevant field equation by variation of an effective action with respect
to $a$
\begin{equation}\label{C9}
\Gamma[a]=2\delta\int d^4x~a^3\left\{-\frac{R}{2}+
\tilde{\zeta}Y_{2,1}+\tilde{\xi}Y_{2,2}\right\}+\Delta\Gamma.
\end{equation}
Here $\Delta\Gamma$ accounts for an energy-momentum tensor of  matter $T_{\mu\nu}$ which has the
usual coupling to the metric. From $\delta\Delta\Gamma/\delta
a=(\delta\Delta\Gamma/\delta g^{ij})
(\delta g^{ij}/\delta a)=(\sqrt{g}g_{ij}p/2)(-2a^{-3}\delta_{ij})=-3pa^2$ one finds formally
\begin{equation}\label{C10}
\Delta\Gamma=-\int d^4 x~p~a^3.
\end{equation}
By partial integration one has
\begin{equation}\label{C11}
\Gamma[a]=2\delta\int d^4x\left\{a\dot{a}^2
\left(3-\frac{3}{2}\tilde{\zeta}-\frac{9}{2}\tilde{\xi}\right)
-\frac{p}{2\delta}a^3\right\}
\end{equation}
and we infer the field equation from $\delta(\Gamma+\Delta\Gamma)/\delta a=0$, i.e.
\begin{equation}\label{C12}
-(3\dot{a}^2+6a\ddot{a})\left(1-\frac{1}{2}\tilde{\zeta}-\frac{3}{2}\tilde{\xi}\right)
=\frac{3}{2\delta}pa^2
\end{equation}
or
\begin{equation}\label{C13}
2\dot{H}+3H^2=-\left(1-\frac{1}{2}\tilde{\zeta}-\frac{3}{2}\tilde{\xi}\right)^{-1}
\frac{p}{2\delta}.
\end{equation}
Combining eq. (\ref{C13})
with energy momentum conservation, $\dot{\rho}+3H(\rho+p)$, this yields the
standard Friedmann cosmology. Of course, these equations can also be derived by inserting
the ansatz (\ref{C1}) into the field equation (\ref{S11D}), cf. appendix A.

The only difference from Einstein gravity turns out to be the
different value of the Planck mass which can be extracted from cosmological observations
as compared to the one inferred
from local gravity measurements. Denoting the reduced Planck mass for cosmological
observations by
\begin{equation}\label{C14}
\bar{M}^2_c=2\delta-\frac{1}{2}\zeta-\frac{3}{2}\xi=\left(1-\frac{3}{2}\beta_A\right)\mu
\end{equation}
and comparing with Newtonian gravity (\ref{LL26}) we infer the ratio
\begin{equation}\label{C15}
\frac{\bar{M}^2_c}{\bar{M}^2}=1-2\beta_A.
\end{equation}
This affects quantitative cosmology like nucleosynthesis or the CMB. In view of the
severe bound on $\beta_A$ from post-Newtonian gravity derived in the preceeding section
these effects are very small, however. For $\beta_A=0$ we recover precisely the standard
Friedmann cosmology. Of course, this could be modified by a cosmological constant or other
degrees of freedom not contained in $E^m_\mu$. In particular, we note that the classical
action (\ref{7}) is dilatation invariant whereas the quantum effects induce a dilatation
anomaly. For a suitable form of the anomaly this could lead to quintessence \cite{Q},
\cite{CQ}.

\section{Partial bosonization}
\label{partialbosonization}

Our aim is a computation of $\Gamma[E^m_\mu]$ for the action (\ref{7}).
An explicit solution of the functional integral (\ref{12}) seems
out of reach and we have to proceed to approximations.
There is no obvious small parameter in the problem since the parameter
$\alpha$ can be rescaled to an arbitrary value by a rescaling of $\psi$. Non-perturbative
approximations will be hard to control but they should give at least an insight into
the qualitative structure.

A convenient tool in our context is partial bosonization. This method reformulates the
fermionic functional integral  (\ref{12}) into an equivalent functional integral
involving both bosonic and fermionic degrees of freedom. In this formulation the
dynamical role of the fluctuations in the ``gravitational degrees of freedom'' will
become apparent. The reformulation is achieved \cite{HS,NJL,GN,EW} by use of
a functional integral over fermions
$\psi$ and bosons $\hat{\chi}^m_\mu$, i.e.
\begin{eqnarray}\label{PB2}
W^{\prime}[J]&=&\ln\int D\psi D\hat{\chi}^n_\nu\exp\left\{\int d^dx\left[
\alpha\det(\tilde{E}^m_\mu-\hat{\chi}^m_\mu)\right.\right.\nonumber\\
&&\left.\left.-\alpha\det(\tilde{E}^m_\mu)
+J^\mu_m\hat{\chi}^m_\mu\right]\right\}.
\end{eqnarray}

We show in appendix B that the free energy $W$ is equivalent to the one of
the original theory (\ref{12}) up to a local polynomial in $\det(J^\mu_m)$
\begin{equation}\label{17AA}
W^\prime[J]=W[J]+\int d^dxF\Big(\det J^\mu_m(x)\Big).
\end{equation}
Performing derivatives of $W$ with respect to $J$ at $J=0$ one obtains the connected
correlation functions for composite fermion bilinears $\tilde{E}^m_\mu$. We see that all
connected correlation functions involving less than $d$ powers of $\tilde{E}^m_\mu$ are
equal for the new ``partially bosonized'' functional integral and the original theory.
(The fermionic correlation functions are equal anyhow.) In particular, the expectation value
$\langle\tilde{E}^m_\mu\rangle$ and the two point function can equally well be
computed in the partially bosonized setting. The first difference appears  in the connected
correlation function for $d$ powers of $\tilde{E}^m_\mu$.
These differences in the high order correlation functions
are not relevant for our discussion and we will
omit the prime on $W$ from now on, treating the partially bosonized theory as an equivalent
version of the original fermionic theory.

It is apparent from eqs. (\ref{13}) (\ref{PB2}) that the expectation
value of $\hat{\chi}^m_\mu$
is given by the vielbein or fermion bilinear
\begin{equation}\label{PB3}
\langle\hat{\chi}^m_\mu\rangle=E^m_\mu=\frac{i}{2}
\langle\bar{\psi}\gamma^m\partial_\mu\psi
-\partial_\mu\bar{\psi}\gamma^m\psi\rangle.
\end{equation}
Using the definition of the effective action (\ref{14}) and performing a variable
shift $\hat{\chi}^m_\mu=E^m_\mu+\chi^m_\mu$ one obtains a convenient implicit functional
integral expression
\begin{eqnarray}\label{PB4}
\Gamma[E^m_\mu]=-\ln&&\int D\psi D\chi^n_\nu\nonumber\\
\exp\Big\{&&\int d^dx\Big[\alpha\det
(\tilde{E}^m_\mu-E^m_\mu-\chi^m_\mu)\nonumber\\
&&-\alpha\det(\tilde{E}^m_\mu)
+J^\mu_m\chi^m_\mu\Big]\Big\}
\end{eqnarray}
where $J^\mu_m$ is given by eq. (\ref{15}).

The classical approximation to $\Gamma$ neglects all fluctuation effects and simply reads
\begin{equation}\label{PB5}
\Gamma_{cl}=\tilde{\alpha}\int d^dxE~,~E=\det(E^m_\mu).
\end{equation}
One easily infers the classical field equation
(\ref{15})
\begin{equation}\label{PB6}
\frac{\tilde{\alpha}}{(d-1)!}\epsilon^{\mu_1\dots\mu_d}\epsilon_{m_1\dots m_d}
E^{m_2}_{\mu_2}\cdots E^{m_d}_{\mu_d}=J^{\mu_1}_{m_1}.
\end{equation}
Whenever $E\neq 0$ we may introduce the inverse vielbein $E^\mu_m$ obeying
the relations (\ref{S5}). For nonzero $E$ the classical field equation therefore becomes
\begin{equation}\label{PB9}
\tilde{\alpha}EE^\mu_m=J^\mu_m.
\end{equation}
We can use this form in order to show that the field equation (\ref{PB9})
has for $J^\mu_m\rightarrow 0$ only solutions with
\begin{equation}\label{PB7}
E=0.
\end{equation}
Indeed, the classical field equation (\ref{PB9})
implies that a nonzero finite value of $E$ is in contradiction with $J^\mu_m=0$.
Of course, $E=0$ does not require $E^m_\mu=0$. For example, a possible solution is
$(\bar{D}<d-1))$
\begin{equation}\label{PB10}
E^m_\mu=\left\{
\begin{array}{ll}
\delta^m_\mu&\textup{for}~~\mu=0\dots\bar{D}~,~m=0\dots\bar{D}\\
0&\textup{otherwise}\\
\end{array} \right.
\end{equation}
For $\bar{D}=3$ this would describe a flat four-dimensional space-time geometry
which we may associate with Minkowski space later. The remaining $d-4$ dimensions
would not admit a metric description, however. We also observe a large degeneracy
of possible classical solutions. Finally, in presence of a nonvanishing energy
momentum tensor (\ref{18AA}) the classical solution reads (for $E\neq 0)$
\begin{equation}\label{32AA}
\tilde{\alpha} g_{\mu\nu}=T_{\mu\nu}.
\end{equation}

\section{Generalized Dirac operator\\
and loop expansion}
\label{generalizeddirac}

This situation is expected to change drastically once the fluctuation effects are
included. A simple approximation includes only the fermionic fluctuations in one loop order.
For this purpose we put $\chi^m_\mu=0$ in eq. (\ref{PB4}) and expand
\begin{equation}\label{PB11}
\det(\tilde{E}^m_\mu-E^m_\mu)=(-1)^dE\big(1-E^\mu_m\tilde{E}^m_\mu+0(\tilde{E}^2)\big).
\end{equation}
This yields the quadratic term in $\psi$,
\begin{equation}\label{PB12}
S_{(2)}=-\frac{i\tilde{\alpha}}{2}\int d^dxE
(\bar{\psi}\gamma^mE^\mu_m\partial_\mu\psi-\partial_\mu\bar{\psi}E^\mu_m\gamma^m\psi)
\end{equation}
and the one loop expression
\begin{eqnarray}\label{PB13}
\Gamma&=&\tilde{\alpha}\int d^dxE+\Gamma_{(1l)}~,\nonumber\\
\Gamma_{(1l)}&=&-\ln\int D\psi\exp\left\{-S_{(2)}[\psi,E^m_\mu]\right\}.
\end{eqnarray}
The Gaussian Grassmann integration for the fermionic one loop contribution can be
evaluated explicitely. We concentrate here on Majorana spinors \footnote{
For Dirac spinors the relevant operator $S^{(2)}_F$ turns out to be the same \cite{CWGG},
but there is no factor $1/2$ in (\ref{PB14}). Also the matrix $C$ is absent. This plays no role
since $\det C=1$.}
where $\bar{\psi}$ and $\psi$ are identified (up to the matrix C)
\begin{equation}\label{PB14}
\Gamma_{(1l)}=-\frac{1}{2}\ln~\det S^{(2)}_F
\end{equation}
with
\begin{equation}\label{PB15}
S^{(2)}_F=-2i\tilde{\alpha}C\left[E\gamma^\mu\partial_\mu
+\frac{1}{2}\partial_\mu(E\gamma^\mu)\right]
\end{equation}
where
\begin{equation}\label{PB16}
\gamma^\mu=E^\mu_m\gamma^m.
\end{equation}

Up to irrelevant constants we can also write
\begin{eqnarray}\label{PB17}
\Gamma_{(1l)}&=&-\frac{1}{2}Tr~\ln(E{\cal D})~,\\
{\cal D}&=&\gamma^\mu\partial_\mu+\frac{1}{2E}\gamma^m
\partial_\mu(EE^\mu_m)=\gamma^\mu\hat{D}_\mu.\label{PB18}
\end{eqnarray}
We call ${\cal D}$ the generalized Dirac operator
and observe the appearance of a ``covariant derivative''
\begin{equation}\label{PB19}
\hat{D}_\mu=\partial_\mu+\frac{1}{2E}E^m_\mu\partial_\nu(EE^\nu_m).
\end{equation}
(For Weyl spinors one should either multiply ${\cal D}$ by an appropriate projection operator
$(1+\bar{\gamma})/2$ or work within a reduced space of spinor indices, using $C\gamma^m$
instead of $\gamma^m$ since only $C\gamma^m$ acts in the reduced space.)
The contribution from the derivative acting on the vielbein can also be written in the form
\begin{equation}\label{33AX}
{\cal D}=\gamma^m(E^\mu_m\partial_\mu-\Omega_m)~,~\Omega_m=-\frac{1}{2E}\partial_\mu
(EE^\mu_m).
\end{equation}

It is instructive to compare the generalized Dirac
operator ${\cal D}$ with the corresponding operator
${\cal D}_E$ in Einstein-gravity. The latter is constructed from the Lorentz covariant
derivative $D_\mu$ which appears in the spinor kinetic term (Majorana spinors)
\begin{eqnarray}\label{PB20}
i\bar{\psi}\gamma^\mu D_\mu\psi&=&i\bar{\psi}\gamma^me^\mu_m
\left(\partial_\mu-\frac{1}{2}\omega_{\mu np}\Sigma^{np}\right)\psi
=i\bar{\psi}{\cal D}_E\psi\nonumber\\
&=&i\bar{\psi}\gamma^\mu\partial_\mu\psi-\frac{i}{4}\Omega_{[mnp]}
\bar{\psi}\gamma^{mnp}_{(3)}\psi.
\end{eqnarray}
Here $\gamma^{mnp}_{(3)}$ is the totally antisymmetrized product of three $\gamma$-matrices
$\gamma^{mnp}_{(3)}=\gamma^{[m}\gamma^n\gamma^{p]}$ and $\Omega_{[mnp]}$ corresponds
to the total antisymmetrization of
\begin{equation}\label{PB21}
\Omega_{mnp}=-\frac{1}{2}e^\mu_me^\nu_n(\partial_\mu e_{\nu p}-\partial_\nu e_{\mu p}).
\end{equation}
Replacing $e^m_\mu$ by $E^m_\mu$ one finds
\begin{equation}\label{PB22}
{\cal D}={\cal D}_E[E]+\frac{1}{4}\Omega_{[mnp]}[E]\gamma^{mnp}_{(3)}.
\end{equation}
For the fermionic loop contribution the only difference between spinor gravity and standard
gravity concerns the totally antisymmetric piece $\sim \Omega_{[mnp]}$!

Neglecting the piece $\sim\Omega_{[mnp]}$ the first contribution ${\cal D}_E[E^m_\mu]$ is
covariant with respect to both general coordinate and {\em local} Lorentz
transformations. Replacing ${\cal D}\rightarrow {\cal D}_E$ in the integral (\ref{PB17}) will
therefore lead to a one loop effective action $\Gamma_{1l}$ with these symmetries.
This is a gravitational effective action of the standard type. Expanded in the
number of derivatives one will find the curvature
scalar plus higher derivative invariants like $R^2,R_{\mu\nu}R^{\mu\nu}$ etc.
However, the additional piece $\sim\Omega_{[mnp]}[E^m_\mu]$ violates the local Lorentz
symmetry and only preserves a global Lorentz symmetry. We therefore expect the
appearance of new terms in the effective action which are invariant under global but
not local Lorentz-rotations. According to eq. (\ref{PB22}) all additional terms must
involve $\Omega_{[mnp]}$ or derivatives thereof. They vanish for $\Omega_{[mnp]}=0$.
As discussed at the end of sect. \ref{linearizedfield} the linear approximation to
$\Omega_{[mnp]}$ only involves $c_{\nu\rho}$ and we can conclude that $\beta_A=0$.
This concerns precisely the ``dangerous term'' restricted by observation
(\ref{R28}). We conclude that one loop spinor gravity is consistent with all
tests of general relativity.

This simple argument can be confirmed by an explicit computation \cite{HW1}. The overall
coefficient $\mu$ in the effective action (\ref{27NNA}), as well as $\epsilon$, depends
on the precise choice of the regularization. In contrast, the relative coefficients
$\tau_A$ and $\beta_A$ are regularization-independent and characterized by the de Witt
coefficients \cite{Wi}, \cite{Tor} of the generalized Dirac operator. One obtains
\cite{GLo,HW1}
\begin{equation}\label{146X}
\tau_A=3~,~\beta_A=0.
\end{equation}

We finally observe that the trace in eq. (\ref{PB17}) involves a trace over spinor
indices as well as an
integration over space coordinates or, equivalently, a momentum integral in Fourier space.
As it stands, these integrations are highly divergent in the ultraviolet and the integral
(\ref{PB17}) needs a suitable regularization. This regularization should preserve the
invariance under general coordinate transformations. If possible, it should also preserve
the global Lorentz symmetry. However, there may be obstructions in the form of
``gravitational anomalies'' \cite{AW} for $d=6~mod~4$. At present it is not known if such
anomalies occur in spinor gravity. For the time being we neglect this possible complication
and assume global Lorentz symmetry of the effective action.

We do not claim quantitative accuracy for our one loop evaluation of the bosonic
effective action. In particular, the value of the coefficient $\tau_A$ may be
affected by higher loop orders. Also dimensional reduction from a higher dimensional
spinor gravity theory will affect the effective four dimensional value of $\tau_A$.
In contrast, our finding $\beta_A=0$ may be more robust. First of all, one may perform
a similar computation \cite{SG2} by the solution of the Schwinger-Dyson equation (without using
partial bosonisation). In lowest order, the nontrivial contribution to $\Gamma$
will again be characterized by eq. (\ref{PB17}) while the coefficient of the
``classical contribution'' (\ref{PB5}) will be modified \cite{BEA}. More generally,
all higher orders in the evaluation of $\Gamma$ using the Schwinger Dyson approach will
involve powers of the exact fermionic propagator in the ``background'' of a vielbein
$E^m_\mu$ \cite{BEA}. One may expand the exact inverse fermionic propagator
$\Gamma^{(2)}_F$ (the generalization of $S^{(2)}_F$ (\ref{PB15}) up to terms
linear and quadratic in $\partial_\mu E^m_\nu$. If the corresponding operator ${\cal D}$
(the generalization of eq. (\ref{PB22})) does not contain the linear field $w^\mu$
(\ref{LL10}) we can conclude that $\beta_A$ vanishes to all orders. First investigations
\cite{SG2} suggest that such a property could be related to a hidden nonlinear
symmetry. As an interesting alternative, $\beta_A=0$ could be associated to an infrared
stable partial fixed point in the flow of generalized couplings.

\section{Conclusions}
\label{conculsions}
In \cite{HW1} and this paper we have formulated a proposal for a
unified theory based only on spinor
fields. We insist on a well defined action which is a polynomial in the spinor
Grassmann field. The action is invariant under general coordinate and global
Lorentz transformations, whereas local Lorentz symmetry may be violated. 
Within spinor gravity the
vielbein is not a fundamental field but rather arises as a
composite object or bound state. It is described by the expectation value of a fermion
bilinear. The metric can be formed as usual from the product of two vielbeins. As a
consequence of the missing local Lorentz symmetry the vielbein contains, however, new
physical degrees of freedom not described by the metric. This leads to a version of
generalized gravity with global instead of local Lorentz symmetry. We discuss in detail
the observational consequences of such a generalization. In particular, we
find that the form suggested by the one loop approximation to spinor gravity is compatible
with all present tests of general relativity. We conclude that the local character of the
Lorentz symmetry is tested only very partially by observations.

Can spinor gravity serve as a candidate for a fundamental theory of all interactions? Several
important steps have to be taken before this question can be answered. First of all, only a well
defined and diffeomorphism invariant regularization procedure for the functional measure
would make the expectation values of fermion bilinears explicitely calculable. Second, the
most general form of the classical action admits many independent
polynomials which are invariant with respect to
diffeomorphisms and global Lorentz rotations. At this stage the corresponding
dimensionless couplings remain undetermined. A predictive unified theory would have to
select a particular point in the high dimensional space of possible couplings. One
possibility is that this particular point is associated to an enhanced symmetry \cite{6AA}. As an
interesting alternative,
the renormalization flow of the couplings could reveal a fixed point
which would render spinor gravity renormalizable. If such a fixed point exists, the number of
relevant (or marginal) parameters at the fixed point would determine the number of free
parameters entering the predictions for physical quantities.
If there is only one relevant direction it could be associated to the overall mass
scale of the theory by dimensional transmutation. In such a case no free dimensionless
coupling would remain and spinor gravity would become completely predictive. Only the number
of dimensions would influence the outcome of a calculation of fermion bilinears like the
vielbein. Third, it remains to be shown that spinor gravity formulated in a suitable
dimension $d>4$ admits an interesting ground state with a small characteristic length scale
for the $d-4$ internal dimensions and a large scale for the observed dimensions.
The isometries of this ground state should be the gauge group
$SU(3)$ x $SU(2)$ x $U(1)$ of the standard model (up to tiny effects of electroweak
symmetry breaking) and the chirality index should account for three generations of quarks and
leptons. Obviously, the way towards such a goal is still long, but, we believe, worthwhile
pursuing.

\vspace{1cm}
\noindent
{\em Acknowledgment}

The author would like to thank to thank A. Hebecker for collaboration on the topics of
several key sections, as summarized in \cite{HW1}. He thanks J. Lukierski for drawing his
attention to the early literature on this subject after the first version of this paper.

\section* {Appendix A: Field equations}
\renewcommand{\theequation}{A.\arabic{equation}}
\setcounter{equation}{0}
In this appendix we give details for the field equations of generalized gravity which are
used in the main text. We use the form (\ref{S11C}) for the effective action.
In order to derive the field equations for the effective action (\ref{S11C}) we
expand the inverse vielbein in linear order
\begin{equation}\label{F.1}
E^\mu_m=\bar{E}^\mu_m+\delta E_m^{\ \mu}.
\end{equation}
For the vielbein, metric and determinant this implies
\begin{eqnarray}\label{F.2}
E^n_\nu&=&\bar{E}^n_\nu-\bar{E}^m_\nu\bar{E}^n_\mu\delta E_m^{\ \mu}\nonumber\\
g^{\mu\nu}&=&\bar{g}^{\mu\nu}+\bar{E}^{m\nu}\delta E_m^{\ \mu}+\bar{E}^{m\mu}
\delta E_m^{\ \nu}~,~\nonumber\\
E&=&\bar{E}(1-\bar{E}^m_\mu\delta E_m^{\ \mu}).
\end{eqnarray}
Omitting from now on the bars, the first variation of the invariants reads
\begin{eqnarray}\label{F.3}
\delta(\epsilon\Gamma_0-\bar{\delta}\Gamma_{2,R})&=&
-\int d^dxE\{\bar{\delta}(2R^m_\mu-RE^m_\mu)\nonumber\\
&&+\epsilon E^m_\mu\}\delta E_m^{\ \mu}
\end{eqnarray}
and
\begin{eqnarray}\label{F.4}
\delta\Gamma_{2,1}&=&\int d^dxE\{4\Omega_{\mu\nu}^{\ \ \ n}
\Omega^{m\nu}_{\ \ \ n}\nonumber\\
&&+2(D_\nu\Omega^{\rho\nu}_{\ \ \ n})E^m_\rho E^n_\mu\nonumber\\
&&-\Omega_{\nu\rho}^{\ \ \ n}
\Omega^{\nu\rho}_{\ \ \ n}
E^m_\mu\}\delta E_m^{\ \mu}~,~\\
\delta\Gamma_{2,2}&=&\int d^dxE\{E^\rho_n\partial_\rho(D^\nu E^n_\nu)
E^m_\mu\nonumber\\
&&-\partial_\mu(D^\nu E^m_\nu)\nonumber\\
&&+\frac{1}{2}D^\nu E^n_\nu D_\rho
E^\rho_n E^m_\mu\}\delta E_m^{\ \mu}.\label{F.5}
\end{eqnarray}
We therefore obtain the field equation from $\delta\Gamma/\delta E_m^{\ \mu}=0$ as
\begin{eqnarray}\label{F.6}
&&\bar{\delta}(2R^m_\mu-RE^m_\mu)+\epsilon E^m_\mu=\qquad~\hspace{2cm}\\
&&\zeta\{4\Omega_{\mu\nu}^{\ \ \ n}\Omega^{m\nu}_{\ \ \ n}+2
(D_\nu\Omega^{\rho\nu}_{\ \ \ n})E^m_\rho E^n_\mu
-\Omega_{\nu\rho}^{\ \ \ n}\Omega^{\nu\rho}_{\ \ \ n}E^m_\mu\}\nonumber\\
&&+\xi\{E^\rho_n\partial_\rho(D^\nu E^n_\nu)E^m_\mu-
\partial_\mu(D^\nu E^m_\nu)+\frac{1}{2}D^\nu
E^n_\nu D_\rho E^\rho_n E^m_\mu\}.\nonumber
\end{eqnarray}
By multiplication with $E_{\nu m}$ we can bring this into the form of a modified
Einstein equation
\begin{eqnarray}\label{F.7}
2\bar{\delta}&&\left(R_{\mu\nu}-\frac{1}{2}R g_{\mu\nu}\right)=-\epsilon g_{\mu\nu}\\
&&+\zeta\{4\Omega_{\mu\rho m}\Omega_\nu^{\ \rho m}-2(D_\rho\Omega^\rho_{\ \nu m})E^m_\mu
-\Omega_{\sigma\rho}^{\ \ \ m}\Omega^{\sigma\rho}_{\ \ \ m}g_{\mu\nu}\}\nonumber\\
&&+\xi\{E^\sigma_m\partial_\sigma(D^\rho E^m_\rho)g_{\mu\nu}-
\partial_\mu(D^\rho E^m_\rho)E_{\nu m}\nonumber\\
&&\qquad+\frac{1}{2}D^\sigma E^m_\sigma D^\rho E_{\rho m} g_{\mu\nu}\}.\nonumber
\end{eqnarray}
This yields the field equation (\ref{S11D}) for generalized gravity.

We next want to specialize to the general static and isotropic ansatz (\ref{RZ1})
of sects. \ref{generalisotropic}, \ref{modificationofthe}.
We first compute the tensor $\Omega_{\mu\nu}^{\ \ \ m}=
-\frac{1}{2}(\partial_\mu E^m_\nu-\partial_\nu E^m_\mu)$. With $\Omega_{0i}^{\ \ \ m}
=\frac{1}{2}\partial_i E^m_0$ one finds
\begin{equation}\label{R6}
\Omega_{0i}^{\ \ \ 0}=\frac{x_i}{2r}f^\prime~,~
\Omega_{0i}^{\ \ \ j}=\frac{h}{2}\delta_{ij}+h^\prime\frac{x_ix_j}{2r}.
\end{equation}
Similarly, we obtain
\begin{equation}\label{R7}
\Omega_{ki}^{\ \ 0}=0~,~\Omega_{ki}^{\ \ j}=-\frac{k}{2}
(\delta_{jk}x_i-\delta_{ji}x_k).
\end{equation}
For the invariant $\Gamma_{2,1}$ we calculate $Y_{2,1}$ which yields eq. (\ref{RZ2}):
\begin{eqnarray}\label{R8}
Y_{2,1}&=&\Omega_{\mu\nu}^{\ \ \ m}\Omega^{\mu\nu}_{\ \ \ m}=g^{ij}\{\Omega_{ik}^{\ \ n}
\Omega_{jl}^{\ \ n}g^{kl}\nonumber\\
&&-2\Omega_{0i}^{\ \ 0}\Omega_{0j}^{\ \ 0}g^{00}+2\Omega_{0i}^{\ \ k}
\Omega_{0j}^{\ \ k}g^{00}\}\nonumber\\
&=&\frac{k^2}{2}\left\{(g^{ij}\delta_{ij})(g^{kl}x_kx_l)-
g^{ij}g^{jk}x_ix_k\right\}\nonumber\\
&&+\frac{h^2}{2}g^{00}(g^{ij}\delta_{ij})\nonumber\\
&&+\frac{1}{2}\left(h^{\prime 2}+\frac{2hh^\prime}{r}-\frac{f^{\prime 2}}{r^2}\right)
g^{00}(g^{kl}x_kx_l)\nonumber\\
&=&\frac{d-2}{2}\left(\frac{k^2r^2}{A}-\frac{h^2}{B}\right)-
\frac{1}{2AB}\left[(h+rh^\prime)^2-f^{\prime 2}\right]\nonumber\\
&=&\frac{d-2}{2r^2}\left(1+\frac{1}{A}-2
\sqrt{\frac{B+r^2h^2}{AB}}\right) \nonumber\\
&&+\frac{1}{8AB(B+r^2h^2)}
\left[B^{\prime 2}+4rB^{\prime}h(h+rh^\prime)\right.\nonumber\\
&&\left.-4B(h+rh^\prime)^2\right].
\end{eqnarray}
Here we use
\begin{equation}\label{R9}
g^{00}=-\frac{1}{B(r)}~,~g^{ij}=\delta_{ij}
-\frac{A(r)-1}{A(r)r^2}x_ix_j
\end{equation}
and
\begin{eqnarray}\label{R10}
g^{ij}\delta_{ij}&=&d-2+\frac{1}{A}~,~g^{kl}x_kx_l=\frac{r^2}{A}~,\nonumber\\
g^{ij}g^{jk}&=&\delta_{ik}-\frac{A^2-1}{A^2r^2}x_ix_k~,\nonumber\\
g^{ij}g^{jk}x_ix_k&=&\frac{r^2}{A^2}.
\end{eqnarray}

In order to evaluate the invariant $\Gamma_{2,2}$ we need
\begin{eqnarray}\label{R11}
\tilde{D}^\mu E^m_\mu&=&g^{\mu\nu}(\partial_\mu E^m_\nu-\Gamma_{\mu\nu}^{\ \ \lambda}
E^m_\lambda)\nonumber\\
&=&\left(\delta_{ij}-\frac{A-1}{Ar^2}x_ix_j\right)
\partial_i E^m_j-\Gamma_\mu^{\ \ \mu i}E^m_i.
\end{eqnarray}
Here we have employed the explicit form of the connection in our cartesian coordinates
\begin{eqnarray}\label{R12}
\Gamma_{00}^{\ \ 0}&=&0~,~\Gamma_{00}^{\ \ i}=\frac{B^\prime}{2rA}x_i~,~
\Gamma_{0i}^{\ \ 0}=\frac{B^\prime}{2rB}x_i~,\\
\Gamma_{ij}^{\ \ 0}&=&0~,~\Gamma_{i0}^{\ \ j}=0~,\nonumber\\
\Gamma_{ij}^{\ \ k}&=&\frac{A-1}{Ar^2}\left(\delta_{ij}x_k-
\frac{x_ix_jx_k}{r^2}\right)+
\frac{A^\prime}{2r^3A}x_ix_jx_k~,\nonumber\\
\Gamma_\mu^{\ \mu 0}&=&0~,\nonumber\\
\Gamma_\mu^{\ \ \mu i}&=&\left\{(d-2)
\frac{A-1}{Ar^2}+\left(\frac{A^\prime}{A}-\frac{B^\prime}{B}\right)
\frac{1}{2rA}\right\}x_i\nonumber.
\end{eqnarray}
This yields
\begin{eqnarray}\label{R13}
D^\mu E^0_\mu&=&\frac{1}{\sqrt{AB}}\big[(d-1)h+rh^\prime\big]~,\\
D^\mu E^k_\mu&=&x_k \left\{\frac{d-2}{r^2}\left(
\sqrt{\frac{B+r^2h^2}{AB}}-1\right)\right.\nonumber\\
&&\left.+\frac{1}{2r\sqrt{AB}\sqrt{B+r^2h^2}}\big[B^\prime+2rh(h+rh^\prime)\big]\right\}
\nonumber
\end{eqnarray}
and we infer the invariant $Y_{2,2}$ in eq. (\ref{R14}).

We can also compute the components of the tensor $\hat{T}_{\mu\nu}$ in the field
equation (\ref{S11D}), (\ref{XF1}),
\begin{equation}\label{F8}
\hat{T}_{\mu\nu}=\zeta \hat{T}^{(1)}_{\mu\nu}+\xi\hat{T}^{(2)}_{\mu\nu}
\end{equation}
One finds for $h=0$ and $d=4$
\begin{equation}\label{F9}
\hat{T}^{(1)}_{00}=-\frac{B^\prime}{Ar}+\frac{A^\prime B^\prime}{4A^2}
+\frac{B^{\prime 2}}{8AB}-\frac{B^{\prime\prime}}{2A}+\frac{B}{r^2}
\left(1-\frac{1}{\sqrt{A}}\right)^2.
\end{equation}
Combining this with the corresponding expressions for
$\hat{T}^{(1)}_{ij}~,~\hat{T}^{(2)}_{00}~ \textup{and}~
\hat{T}^{(2)}_{ij}$
one may compute the isotropic field equations in a formally more direct but computationally
more cumbersome way as compared to
sect. \ref{modificationofthe}. For this purpose we also need
\begin{eqnarray}\label{F13}
R_{00}&=&\frac{d-2}{2}\frac{B^\prime}{rA}+\frac{B^{\prime\prime}}{2A}-
\frac{A^\prime B^\prime}{4A^2}-\frac{B^{\prime 2}}{4AB}~,~\nonumber\\
R_{ij}&=&\delta_{ij}\left\{\frac{A^\prime}{2rA^2}-\frac{B^\prime}{2rAB}+
(d-3)\frac{A-1}{Ar^2}\right\}\nonumber\\
&&+\frac{x_ix_j}{r^2}\left\{\frac{B^{\prime 2}}{4B^2}+\frac{B^\prime A^\prime}{4AB}-
\frac{B^{\prime\prime}}{2B}+\frac{B^\prime}{2rAB}\right.\nonumber\\
&&\left.+\left(d-2-\frac{1}{A}\right)\frac{A^\prime}{2rA}-(d-3)\frac{A-1}{Ar^2}\right\}~,
\nonumber\\
R_{0i}&=&0
\end{eqnarray}
and
\begin{eqnarray}\label{F14}
R&=&-\frac{B^{\prime\prime}}{AB}+\frac{B^\prime}{2AB}\left(\frac{A^\prime}{A}
+\frac{B^\prime}{B}\right)+\frac{d-2}{rA}\left(\frac{A^\prime}{A}-\frac{B^\prime}{B}\right)\nonumber\\
&&+(d-3)(d-2)\frac{A-1}{Ar^2}.
\end{eqnarray}

Let us finally turn to the computation of the field equation relevant for cosmology, with
the ansatz (\ref{C1}) for the vielbein. For a computation of $\hat{T}_{\mu\nu}$ we need
the components of $E^m_\mu D_\rho\Omega^\rho_{\ \nu m}$, i.e.
\begin{eqnarray}\label{Z1}
E^m_0D_\rho\Omega^\rho_{\ 0m}&=&0~,~D_\rho\Omega^\rho_{\ 0i}=D_\rho
\Omega^\rho_{\ i0}=0~,\nonumber\\
E^m_jD_\rho\Omega^\rho_{\ im}&=&\frac{1}{2}a^2(\dot{H}+2H^2)\delta_{ij}
\end{eqnarray}
and $\Omega_{\mu\rho m}\Omega_\nu^{\ \ \rho m}$, i.e.
\begin{eqnarray}\label{Z2}
\Omega_{0\rho m}\Omega_0^{\ \rho m}&=&\frac{3}{4}H^2~,~\Omega_{0\rho m}
\Omega_i^{\ \rho m}=0,\nonumber\\
\Omega_{i\rho m}\Omega_j^{\ \rho m}&=&-\frac{1}{4}H^2a^2\delta_{ij}.
\end{eqnarray}
We may also use
\begin{equation}\label{Z3}
E^\sigma_m\partial_\sigma(D^\rho E^m_\rho)=-3\dot{H}.
\end{equation}
This yields
\begin{eqnarray}\label{Z4}
\hat{T}_{00}&=&\frac{3}{2}(\zeta+3\xi)H^2~,\nonumber\\
\hat{T}_{ij}&=&-\frac{1}{2}(\zeta+3\xi)(2\dot{H}+3H^2)g_{ij}.
\end{eqnarray}
With
\begin{eqnarray}\label{Z5}
R_{00}-\frac{1}{2}Rg_{00}&=&3H^2~,\nonumber\\
R_{ij}-\frac{1}{2}Rg_{ij}&=&-(3H^2+2\dot{H})g_{ij}
\end{eqnarray}
one finally finds
\begin{eqnarray}\label{Z6}
(3H^2+2\dot{H})\left(1-\frac{\tilde{\zeta}}{2}
-\frac{3\tilde{\xi}}{2}\right)&=&-\frac{p}{2\delta}~,\nonumber\\
3H^2\left(1-\frac{\tilde{\zeta}}{2}-\frac{3\tilde{\xi}}{2}\right)&=&\frac{\rho}{2\delta}.
\end{eqnarray}
The first equation coincides with eq. (\ref{C13}) whereas the combination of both equations
ensures the energy-momentum conservation of  matter
\begin{equation}\label{Z7}
\dot{\rho}=-3H(\rho+p).
\end{equation}

\section* {Appendix B: Functional identity for partial bosonization}
\renewcommand{\theequation}{B.\arabic{equation}}
\setcounter{equation}{0}
In this appendix we derive the partially bosonized functional integral (\ref{PB2}).
For this purpose we use the identity
(recall $\tilde{E}^m_\mu=i(\bar{\psi}\gamma^m\partial_\mu\psi-\partial_\mu
\bar{\psi}\gamma^m\psi)/2$)
\begin{eqnarray}\label{PB1}
\int D\hat{\chi}^n_\nu \exp&&\int d^dx\{\alpha\det(\tilde{E}^m_\mu-
\hat{\chi}^m_\mu)-J^\mu_m(\tilde{E}^m_\mu-\hat{\chi}^m_\mu)\}\nonumber\\
=\exp &&\int d^dx\tilde{V}
\Big(J^\mu_m(x)\Big)
\end{eqnarray}
with
\begin{equation}\label{16AA}
\tilde{V}(J^\mu_m)=\ln\int d\tilde{\chi}^n_\nu\exp
[-\tilde{\alpha}\det(\tilde{\chi}^m_\mu)+J^\mu_m\tilde{\chi}^m_\mu]
\end{equation}
an even function of $J^\mu_m$ for $d$ even and $\tilde{\alpha}=(-1)^{d+1}\alpha$. (We omit
the irrelevant additive constant for $J=0)$.
Furthermore, $\tilde{V}$ is invariant under global Lorentz-transformations of $J^\mu_m$.
More generally, $\tilde{V}$ remains invariant under all special linear transformations
acting on the $d$ x $d$ matrix $J$ from the left or right. (This follows from the invariance
of the integral (\ref{16AA}) under accompanying (inverse) special linear transformations
acting on the integration variable $\tilde{\chi}$.) Therefore $\tilde{V}$ can only be a
function of the determinant $\det(J^\mu_m)$. The definition of the integral (\ref{16AA})
may be somewhat formal (even after subtraction of the value for $J=0$) since
$\det(\tilde{\chi}^m_\mu)$ has positive and negative eigenvalues. (Note that $\tilde{\alpha}$
is imaginary for a Minkowski signature.) We assume that it can be suitably regularized such
that $\tilde{V}$ is analytic in $J$. We then conclude that $\tilde{V}$ is an analytic
function of $\det(J^\mu_m)$,
\begin{eqnarray}\label{16BB}
\tilde{V}(J^\mu_m)&=&F\Big(\det(J^\mu_m)\Big)\nonumber\\
&=&\beta_1\det(J^\mu_m)+\beta_2\big(\det(J^\mu_m)\Big)^2+\dots
\end{eqnarray}
The relation (\ref{17AA}) is now easily obtained by performing in eq. (\ref{PB2}) the
functional integral over $\hat{\chi}^m_\mu$ using eq. (\ref{PB1}).

\end{document}